\documentclass[twocolumn]{aastex631}%,linenumbers
\usepackage{epsfig, bm, times}%, aas_macros}
\usepackage[english]{babel}
\usepackage{amsmath}
\usepackage{amssymb}
\usepackage{graphicx}
\usepackage{xcolor}

\newcommand{\sect}[1]{Sec. \ref{#1}}
\newcommand{\sects}[1]{Secs. \ref{#1}}

\newcommand{\tab}[1]{Tab. \ref{#1}}
\newcommand{\tabs}[1]{Tabs. \ref{#1}}
\renewcommand{\fig}[1]{Fig. \ref{#1}}
\newcommand{\figs}[1]{Figs. \ref{#1}}
\newcommand{\equ}[1]{Eq. \eqref{#1}}
\newcommand{\me}{\textsc{MESA}}
\newcommand{\kep}{\textsc{Kepler}}
\newcommand{\minb}{\textsc{MINBAR}}

\newcommand{\md}{\dot{m}}
\newcommand{\MD}{\dot{M}}
\newcommand{\medd}{\dot{m}_{\rm{Edd}}}
\newcommand{\Medd}{\dot{M}_{\rm{Edd}}}
\newcommand{\mdstab}{\md_{\rm{stab}}}

\newcommand{\mdt}{\md_{\rm{loc}}}
\newcommand{\mda}{\left<\md\right>}
\newcommand{\mdo}{\mda_{\rm{obs}}}

\newcommand{\mdto}{\md_{\rm{to}}}
\newcommand{\ele}[2]{{^{#2}\rm{#1}}}
\newcommand{\mafra}[1]{X_{\rm{#1}}}
\newcommand{\appr}{\texttt{approx140}}
\newcommand{\Lb}{L_{\rm{b}}}
\newcommand{\teff}{T_{\rm{eff}}}
\newcommand{\rb}{R_{\rm{b}}}
\newcommand{\tr}{t_{\rm{rec}}}
\newcommand{\ave}[1]{\bar{#1}}

\newcommand{\mfr}{\epsilon}
\newcommand{\Alo}{\Delta A_{\rm{local}}}
\newcommand{\figscheme}{
\begin{figure*}
  \centering
  \includegraphics[width=0.90\textwidth]{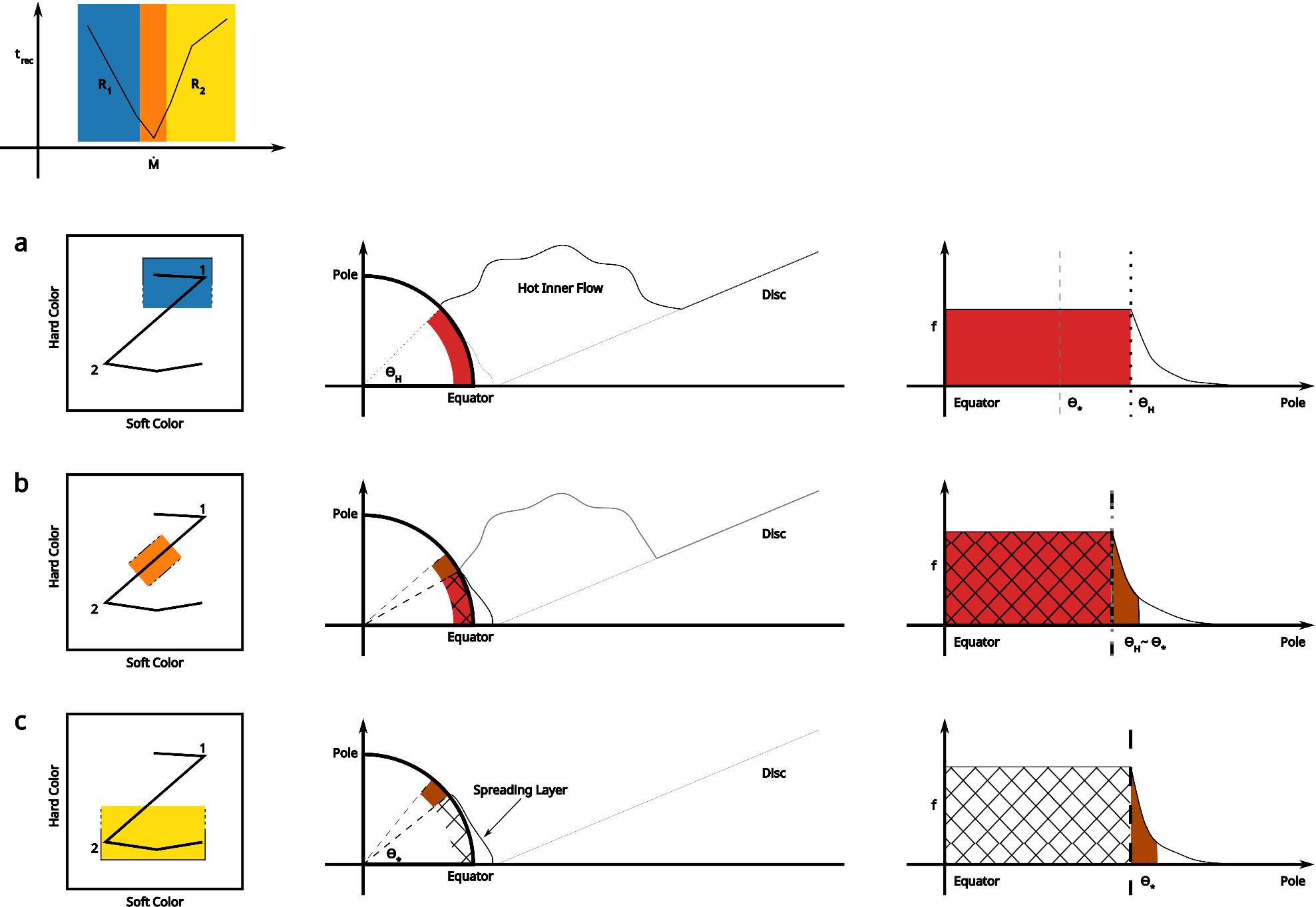}
  \caption{Schematic of the solution proposed in this paper. \textbf{Top panel:} evolution of the burst recurrence time vs. mass accretion rate. The different background colors correspond to the spectral states: hard state (blue), intermediate transition (orange), and soft state (yellow). The lower three rows correspond to each of these states. \textbf{Left column:} position of the source in the spectral color-color diagram. \textbf{Central column:} accretion configuration. We indicate the neutron star, the thin disk, the hot inner flow, and the spreading layer. \textbf{Right column:} corresponding distribution of the matter on the surface, through the parameter $f(\theta)$.  \textbf{First row:} R1, the thin  disk is truncated, and an inner flow reaches the star. The spreading layer (indicated in light gray) does not develop. The burst behavior is determined by a band around the equator set by the hot inner flow (in red, up to the latitude of the hot inner flow $\theta_{\rm{H}}$). The recurrence time decreases with $\MD$ as expected. The efficiency (not shown) is high, indicating that no pollution is present. \textbf{Second row:} the inner flow has almost disappeared, and the spreading layer emerges (up to the latitude $\theta_*$, now on the order of $\theta_{\rm{H}}$). The burning around the equator is transitioning to stable burning (red and hatched). After the turnover the burning is determined by the regions outside the stable zone (brown). The burst recurrence time turns over and begins to increase, and the efficiency decreases. \textbf{Third row:} R2, the accretion is now dominated by the spreading layer which is growing closer to the pole as a function of accretion rate. The lower latitudes are stable, and the burning is only determined by the unstable region closer to the pole. The burst recurrence time keeps increasing and the efficiency decreasing due to the large amount of ashes present. \emph{Angles and sizes are not to scale. The rectangles on the color-color diagrams have dashed boundaries to indicate that the regions are indicative and the precise limits vary from source to source.}}
  \label{fig:scheme}
\end{figure*}
}
\newcommand{\resufigB}{
  \begin{figure*}
    \centering
    \includegraphics[width=0.90\textwidth]{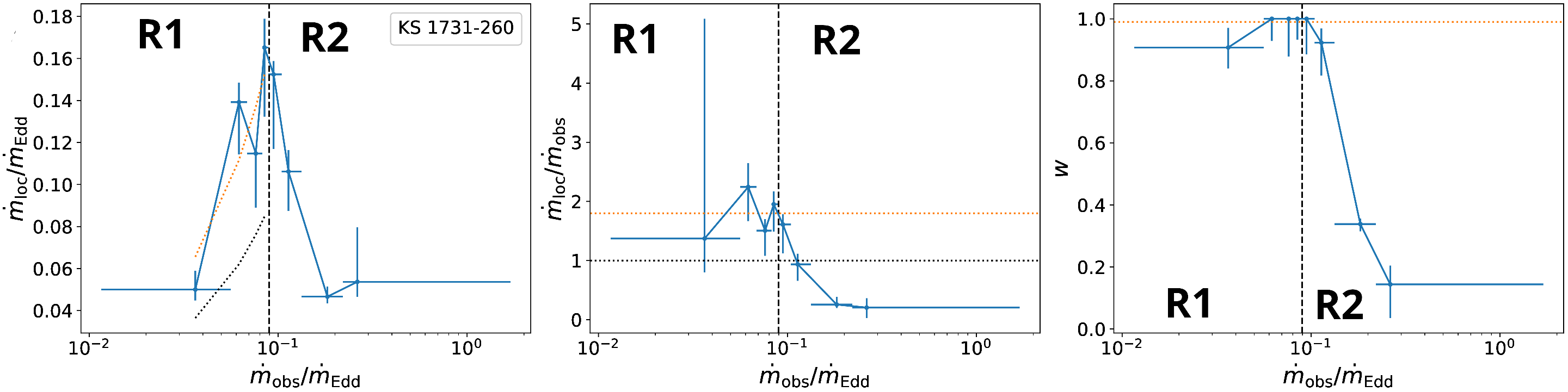}\\
    \includegraphics[width=0.90\textwidth]{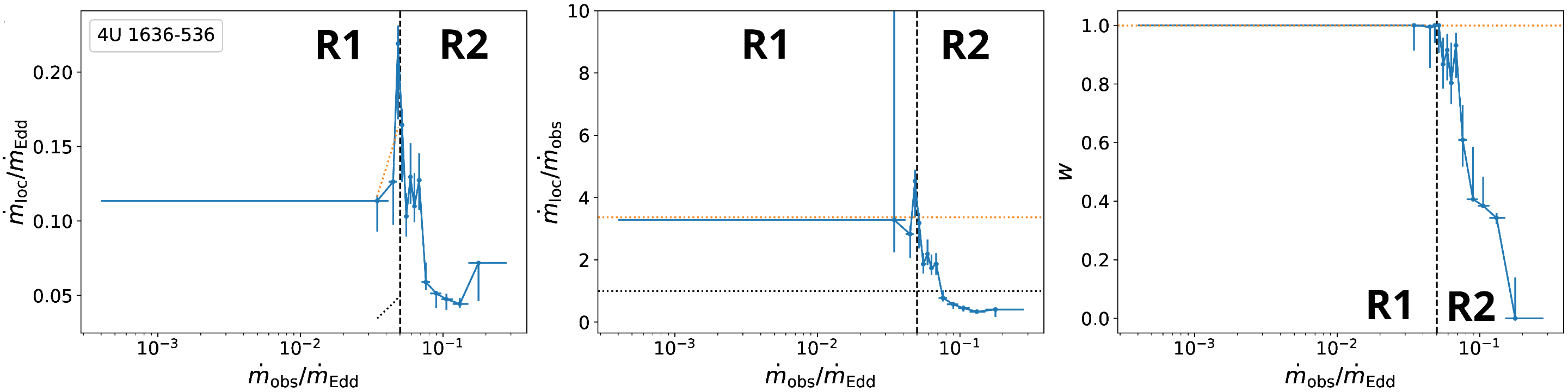}\\
    \includegraphics[width=0.90\textwidth]{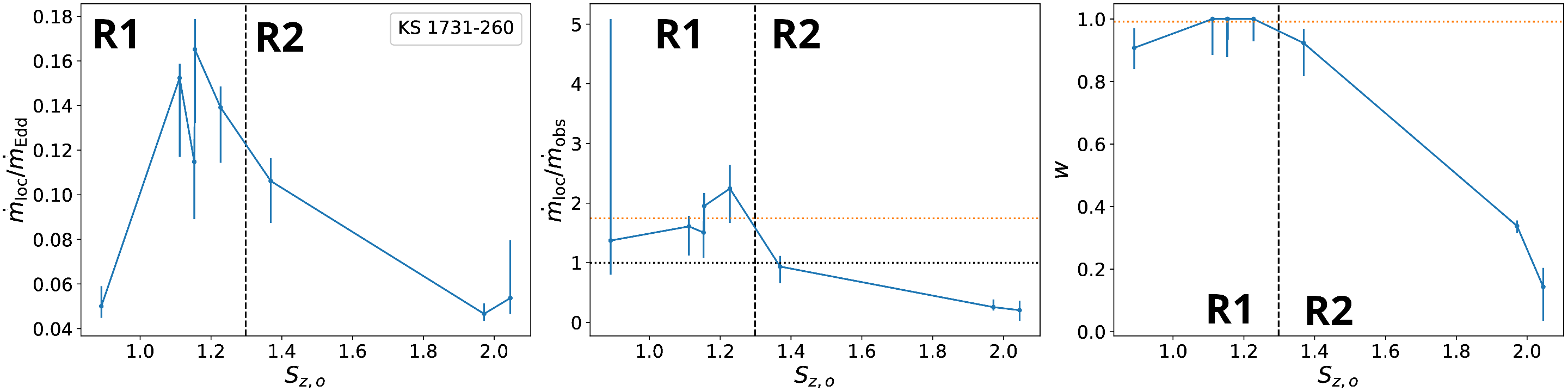}\\
    \includegraphics[width=0.90\textwidth]{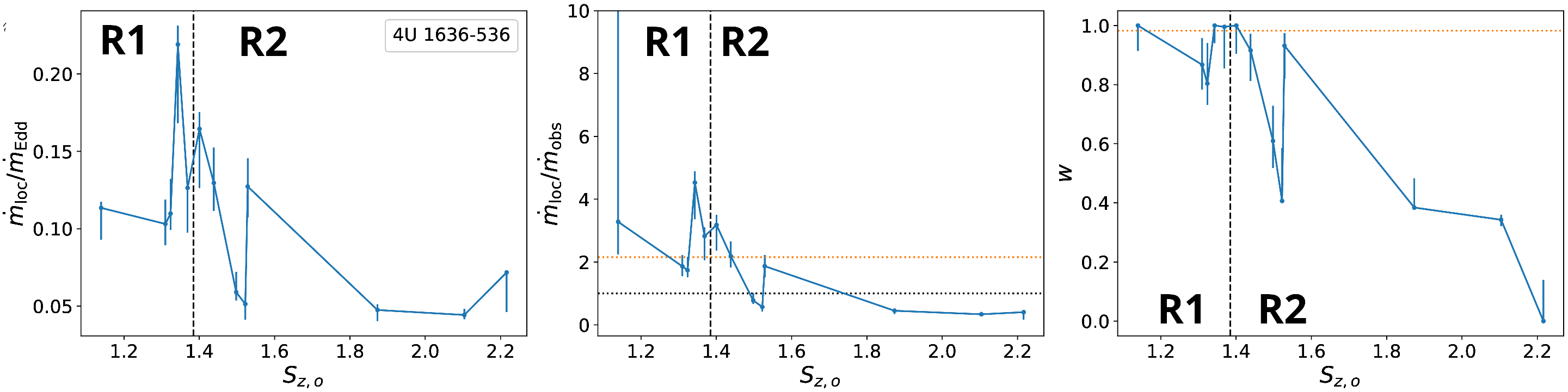}\\
    \caption{Results of the fit of the burst database to the observations for two of the five sources considered in this paper. \textbf{Top two rows:} Each quantity is plotted as a function of $\mdo$ in units of $\medd$. \textbf{Bottom two rows:} Each function is plotted as a function of $S_z$, which tracks the position on the spectral color-color diagram for each source. The vertical (black, dashed) lines in each panel indicate the approximate position of the turnover in $\mdo$ and $S_z$. \textbf{Left column:} fit local $\mdt$ in units of $\medd$. The black dotted line shows the values expected if $\mdt$ and $\mdo$ coincided, while the orange line assumes $\mdt = \bar{f} \mdo$, with $\bar{f}$ set by the average value of the ratio during the regime R1. \textbf{Central column:} $f = \mdt / \mdo$. The horizontal orange line indicates the average value during regime R1. The black and orange lines in the left and middle columns correspond to each other. \textbf{Right column:} the purity $w$ as a function of $\mdo$. The horizontal orange line indicates the average value during regime R1. The values of the turnover $\mdto$ and $S_{z,\, \rm{to}}$ and the averages are reported in \tab{tab:res}.}
    \label{fig:res}
  \end{figure*}
}
\newcommand{\resufigM}{
  \begin{figure*}
    \centering
    \includegraphics[width=0.80\textwidth]{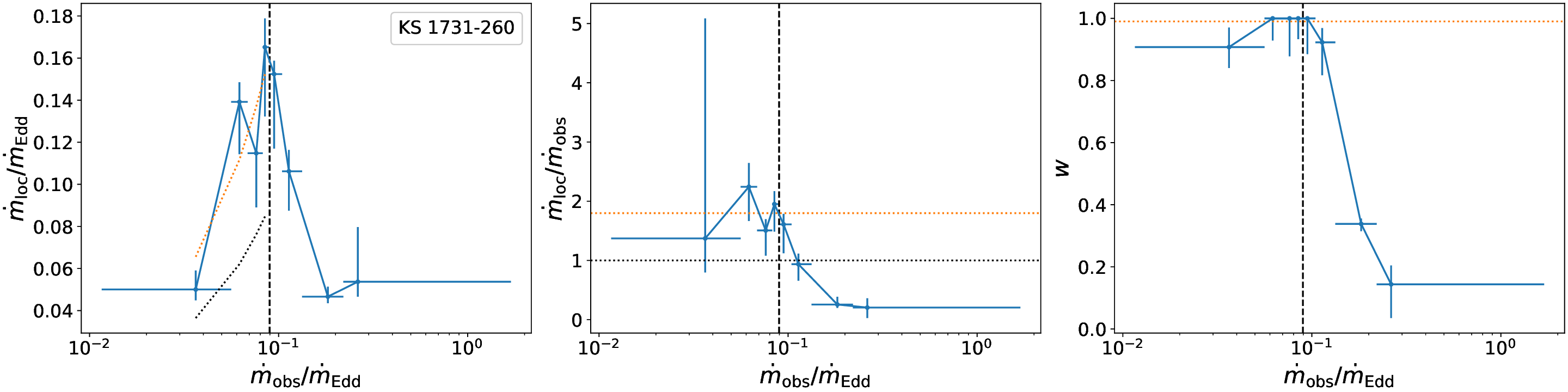}\\
    \includegraphics[width=0.80\textwidth]{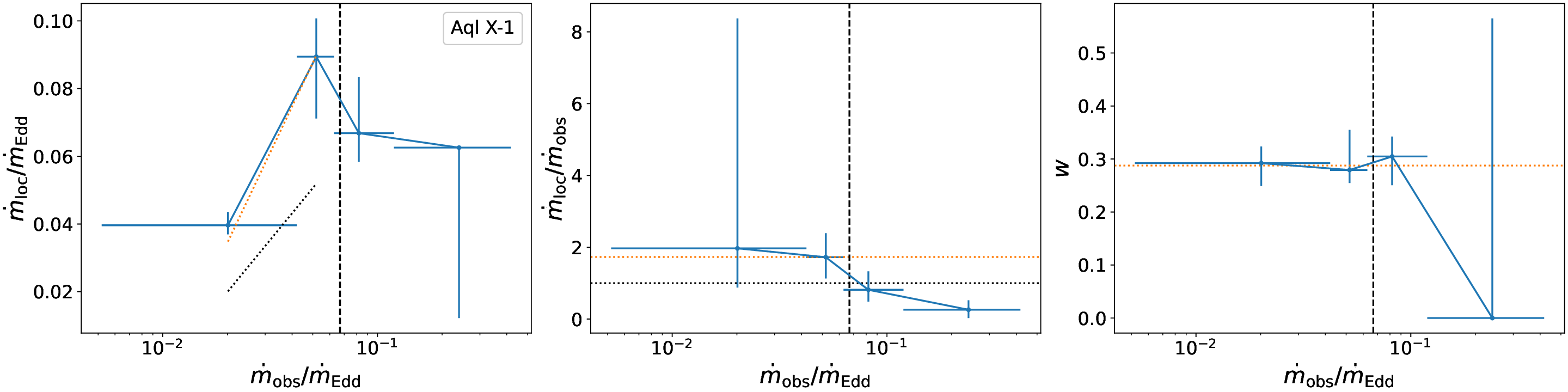}\\
    \includegraphics[width=0.80\textwidth]{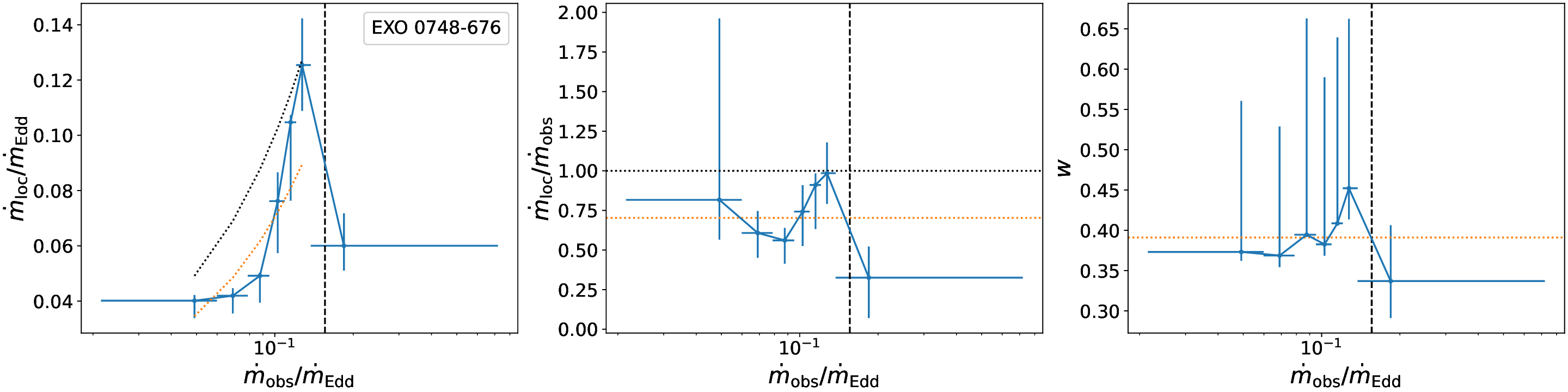}\\
    \includegraphics[width=0.80\textwidth]{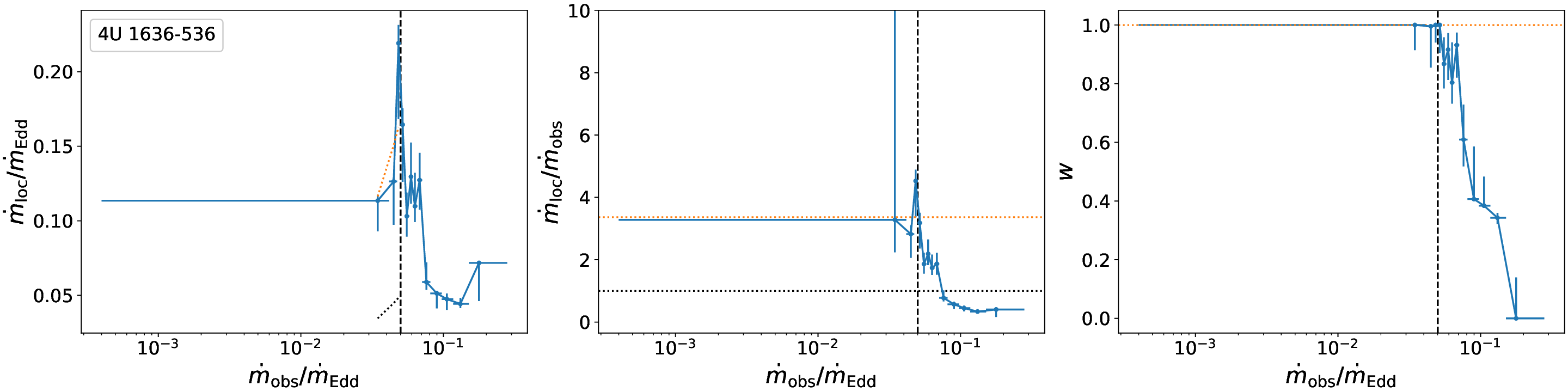}\\
    \includegraphics[width=0.80\textwidth]{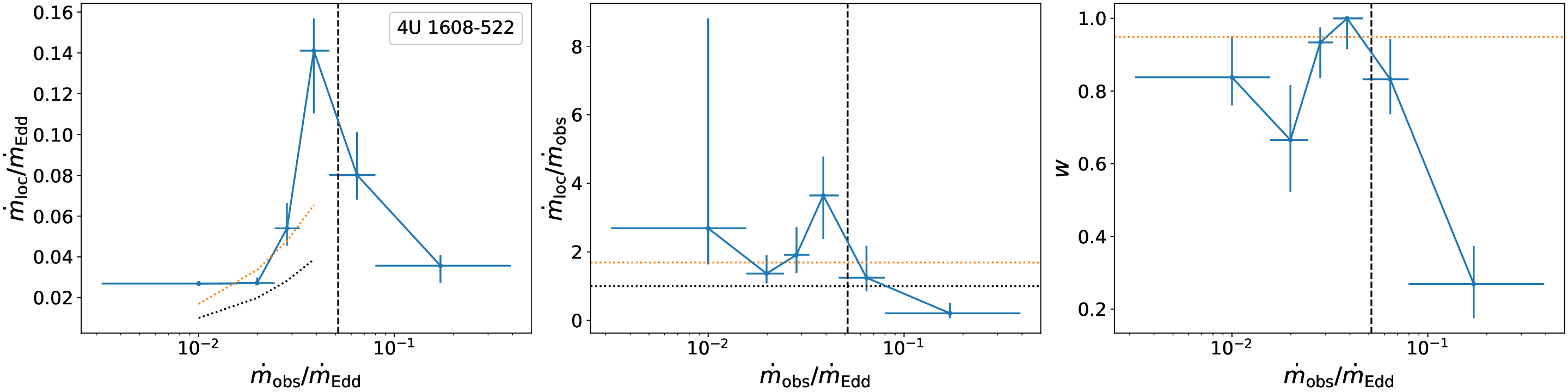}
    \caption{Results of the fit of the burst database to the observations of the 5 sources considered in this paper. Data are displayed as a function of $\mdo/\medd$. See \fig{fig:res} for details and \tabs{tab:fitres} and \ref{tab:res}.}
    \label{fig:resM}
  \end{figure*}
}
\newcommand{\resufigS}{
  \begin{figure*}
    \centering
    \includegraphics[width=0.80\textwidth]{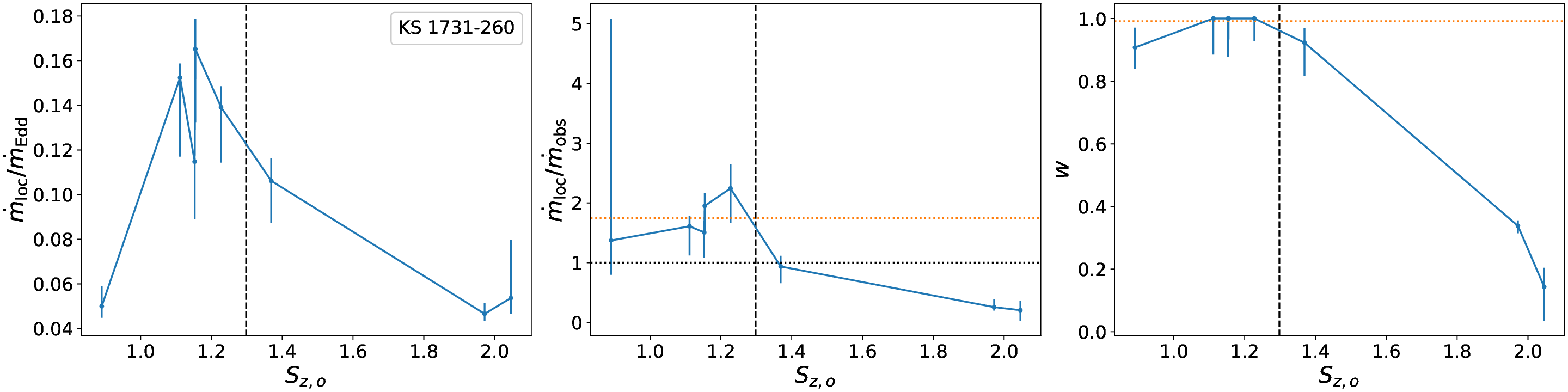}\\
    \includegraphics[width=0.80\textwidth]{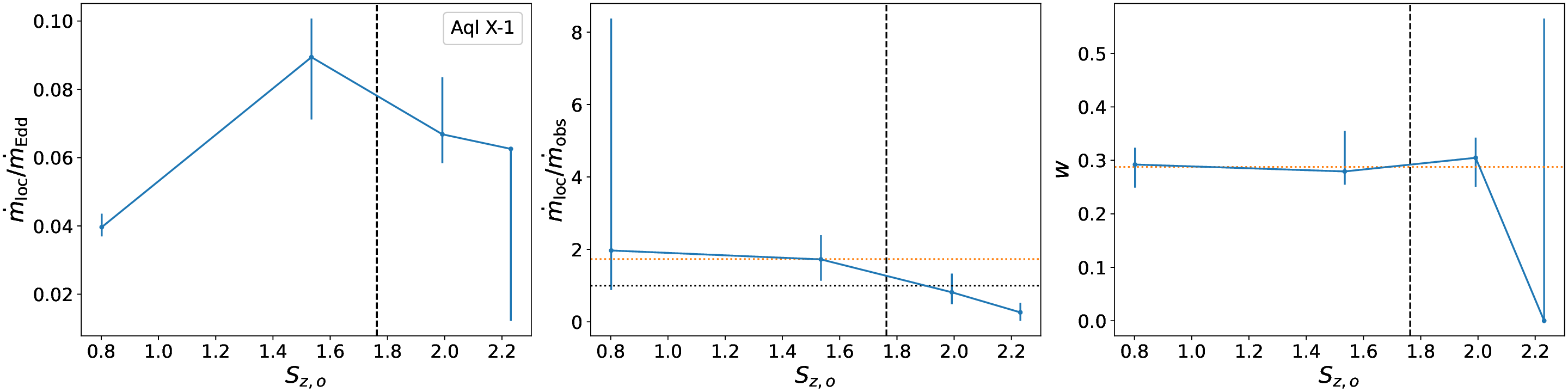}\\
    \includegraphics[width=0.80\textwidth]{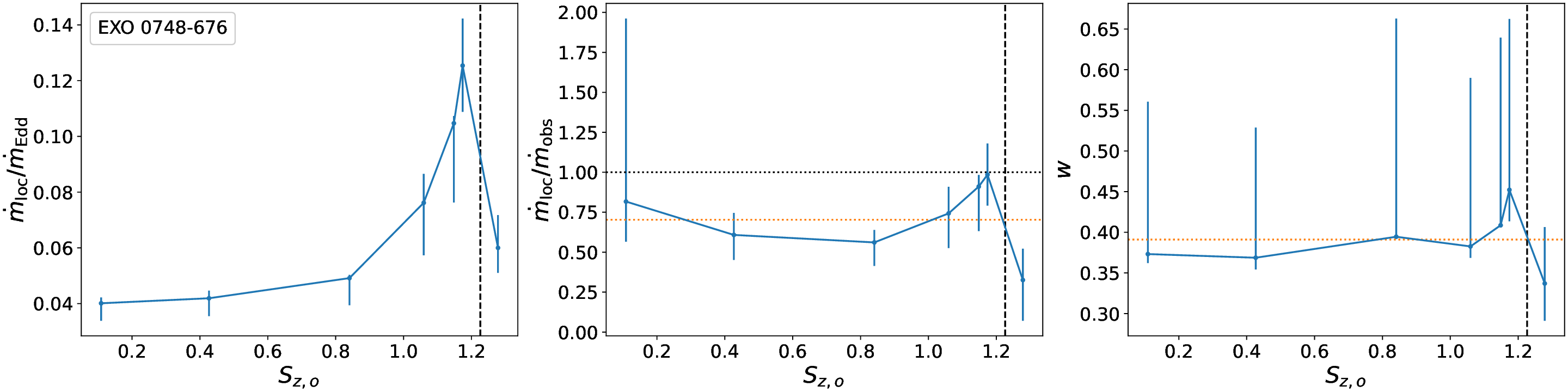}\\
    \includegraphics[width=0.80\textwidth]{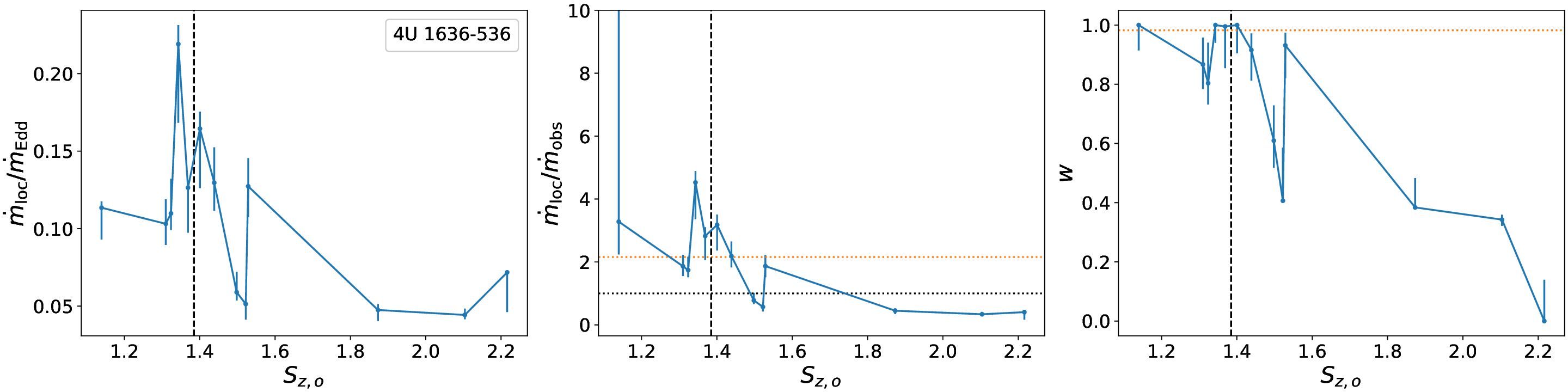}\\
    \includegraphics[width=0.80\textwidth]{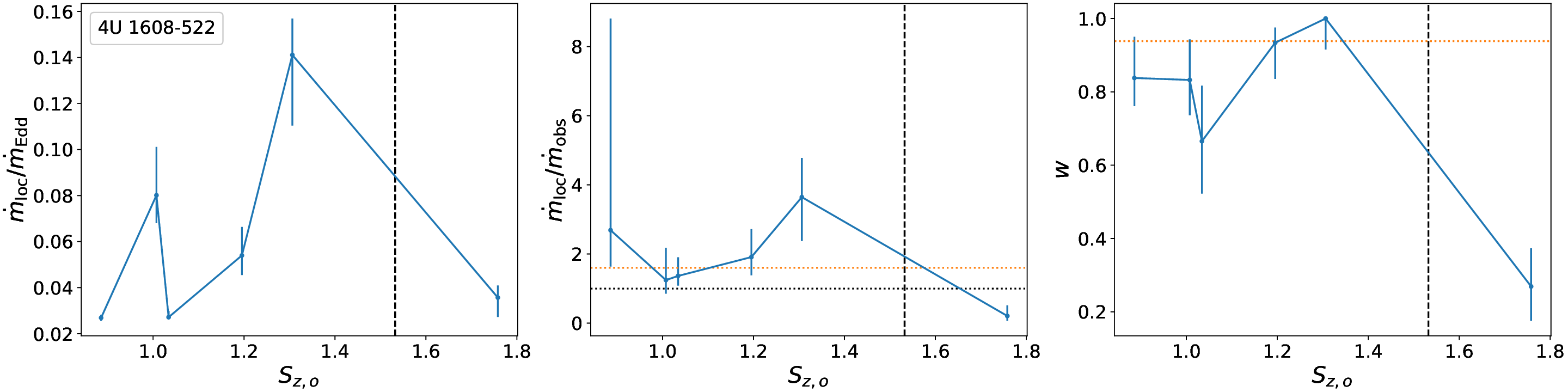}
    \caption{Same as \fig{fig:resM}, but displayed as a function of $S_z$.}
    \label{fig:resS}
  \end{figure*}
}
\newcommand{\figmax}{
\begin{figure}
  \centering
  \includegraphics[width=0.45\textwidth]{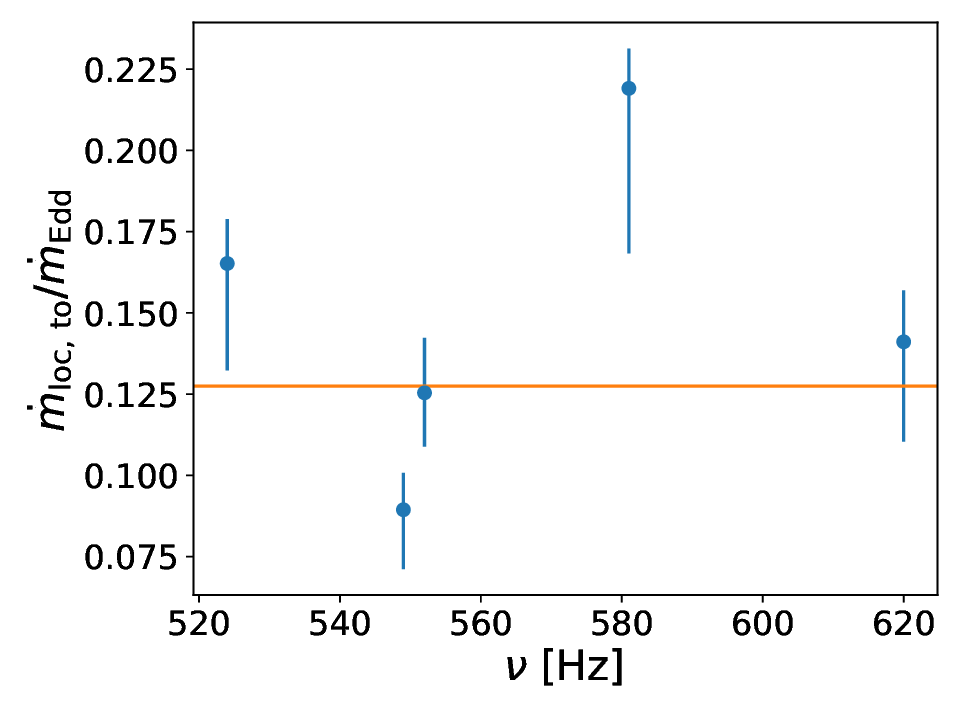}
  \caption{Turnover local accretion rates. Errors correspond to 68\% confidence intervals. The orange line is the weighted average of the values (0.13).}
  \label{fig:peaks}
\end{figure}
}
\newcommand{\figudb}{
  \begin{figure*}
    \centering
    \includegraphics[width=0.45\textwidth]{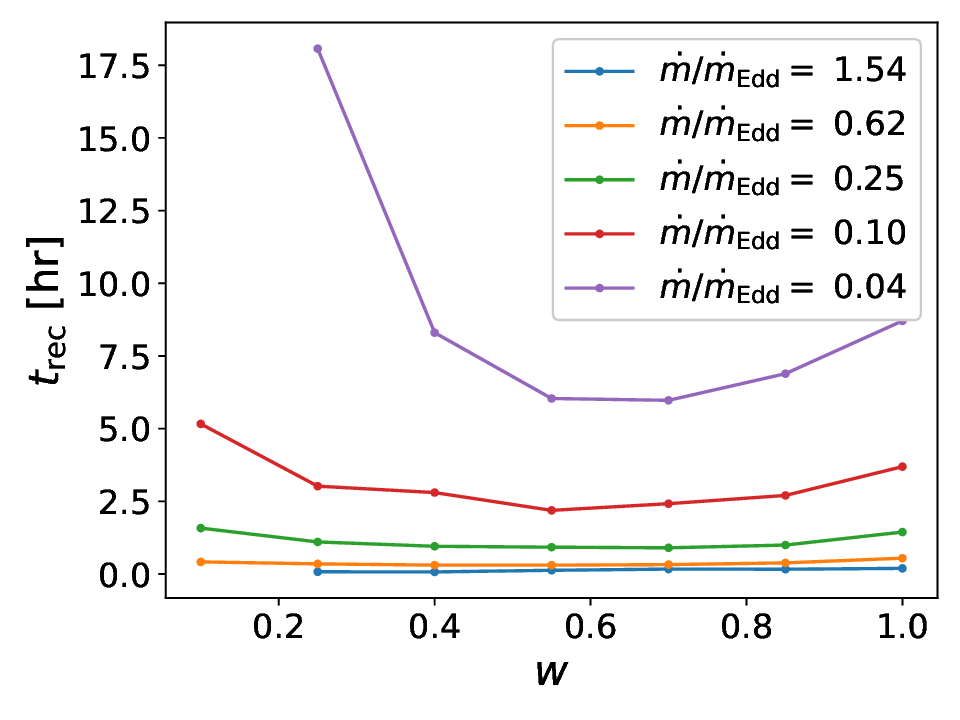}
    \hspace{\fill}
    \includegraphics[width=0.45\textwidth]{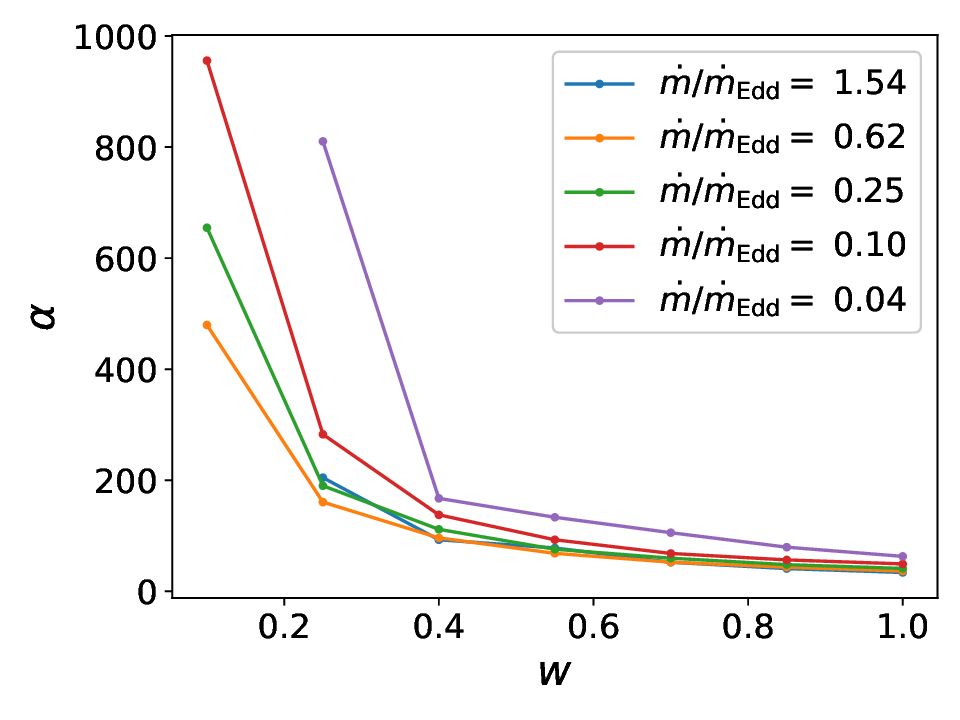}
    \caption{Burst recurrence time $t_{\rm{rec}}$ (left) and $\alpha$ (right) for the database models. The results are displayed as a function of weight $w$ and accretion rate $\mdt / \medd$. The missing points are due to burst quenching.}
    \label{fig:db}
  \end{figure*}
}
\begin{document}

\title{A solution to the tension of burning on neutron stars and nuclear physics}

\author[0000-0002-6447-3603]{Y. Cavecchi}
\affil{Departament de Fis\'{i}ca, EEBE, Universitat Polit\`ecnica de Catalunya, Av. Eduard Maristany 16, 08019 Barcelona, Spain}
\affil{Center for Nuclear Astrophysics across Messengers (CeNAM), 640 S Shaw Lane, East Lansing, MI 48824, USA}
\author[0000-0002-6558-5121]{D. K. Galloway}
\affil{School of Physics and Astronomy, 19 Rainforest Walk, Monash University, Victoria 3800, Australia}
\affil{Institute for Globally Distributed Open Research and Education (IGDORE)}
\affil{Center for Nuclear Astrophysics across Messengers (CeNAM), 640 S Shaw Lane, East Lansing, MI 48824, USA}
\author[0000-0002-3684-1325]{A. Heger}
\affil{School of Physics and Astronomy, 19 Rainforest Walk, Monash University, Victoria 3800, Australia}
\affil{Argelander-Institut f{\"u}r Astronomie, Auf dem H{\"u}gel 71, 53121 Bonn, Germany}
\affil{Center for Nuclear Astrophysics across Messengers (CeNAM), 640 S Shaw Lane, East Lansing, MI 48824, USA}
\author[0009-0002-6591-9870]{P. Santill\'an-Ortega}
\affil{Instituto de Astronom\'ia, Universidad Nacional Aut\'onoma de M\'exico, Ciudad de M\'exico, CDMX 04510, Mexico}
\author[0000-0003-2334-6947]{M. Nava-Callejas}
\affil{Institut d'Astronomie et d'Astrophysique, CP-226, Universit\'e Libre de Bruxelles, 1050 Brussels, Belgium}
\affil{Instituto de Astronom\'ia, Universidad Nacional Aut\'onoma de M\'exico, Ciudad de M\'exico, CDMX 04510, Mexico}
\author[0000-0002-1481-1870]{F. M. Vincentelli}
\affil{Fluid and Complex Systems Centre, Coventry University, Coventry, V1 5FB, UK}
\affil{INAF-Istituto di Astrofisica e Planetologia Spaziali, Via Fosso del Cavaliere, 100 - I-00133 Rome, Italy}
\affil{School of Physics \& Astronomy, University of Southampton, Southampton
SO17 1BJ, UK}
\author[0000-0002-9396-7215]{L. E. Rivera Sandoval}
\affil{Dept. of Physics and Astronomy, University of Texas Rio Grande Valley, Brownsville, TX 78520, USA}
\author[0000-0003-3441-8299]{A. Goodwin}
\affil{International Centre for Radio Astronomy Research - Curtin University, GPO Box U1987, Perth, WA 6845, Australia}
\author[0000-0003-4023-4488]{Z. Johnston}
\affil{Department of Physics and Astronomy, Michigan State University, East Lansing, MI 48824, USA}
\author{S. Puente Mancilla}
\affil{Instituto de Astronom\'ia, Universidad Nacional Aut\'onoma de M\'exico, Ciudad de M\'exico, CDMX 04510, Mexico}
\author[0000-0003-2498-4326]{D. Page}
\affil{Instituto de Astronom\'ia, Universidad Nacional Aut\'onoma de M\'exico, Ciudad de M\'exico, CDMX 04510, Mexico}

\correspondingauthor{Y. Cavecchi}
\email{yuri.cavecchi@upc.edu}

\begin{abstract}
  When neutron stars  accrete matter from a companion star, this matter forms a disk around them and eventually falls on their surface.   Here, the fuel can ignite into bright flashes called Type I bursts.  Theoretical calculations based on state-of-the-art nuclear reactions are able to explain many features of the bursts. However, models predict that the bursts should cease at high accretion rates, whereas in many sources they disappear at much lower rates.  Moreover, their recurrence times also show strong discrepancies with predictions.
  Although various solutions have been proposed, none can account for all the observational constraints.
  Here, we describe a new model that explains all the contradictory behaviors within a single picture.  We are able to reconstruct the conditions on the star surface that determine the burst properties by comparing data to new simulations.  We find strong evidence that the physical mechanism driving the burst behavior is the structure of the accretion disk in the regions closest to the star.  This connection reconciles the puzzling burst phenomenology with nuclear physics and also opens a new window on the study of accretion processes around compact objects.
\end{abstract}


\section{Introduction}
\label{sec:intro}

Type I X-ray bursts are thermonuclear explosions on the surface of accreting neutron stars \citep[see][for a comprehensive review]{rev-arx-2017-2021-gall-keek}. They burn the material that is accreted from a companion star in a low-mass X-ray double star system.  Such systems harbor low-mass ($\lesssim1\ M_\odot$) donor stars, so that the accreted material is expected to be rich in hydrogen and helium, plus some traces of CNO elements. This material is buried in the outer envelope of the neutron star while being compressed by the continuous accretion flow. Some of the material burns stably while sinking, but under certain ranges of accretion rate, $\MD$, the freshly accreted matter reaches a depth where the cooling processes cannot counterbalance the heating due to the nuclear burning and compressional heating. This condition leads to unstable burning that triggers a thermonuclear runaway, which is observed as a sudden rise in the X-ray emission of the neutron star \citep[e.g.,][]{art-1981-fuji-han-miy,rev-arx-2017-2021-gall-keek}.

Numerical models of thermonuclear bursts, coupled to networks of  hundreds of nuclear species, have demonstrated that the burst properties depend on the {\it local} mass accretion rate $\md$ (mass per unit time and surface area), not the {\it global} rate $\MD=4\pi R^2\mda$ (where the average is taken over the entire neutron star surface).  Given the expected solar composition, the bursts should evolve through different ignition regimes as the accretion rate increases, until the burning becomes stable around the Eddington accretion rate \citep[$\mdstab \sim \medd\approx8.8\times10^4\ {\rm g\,cm^{-2}\,s^{-1}}$;][]{art-1981-fuji-han-miy,rev-2003-2006-stro-bild,art-2016-kee-heg}.  The observed accretion rate at which many sources stop bursting is, however, $0.1 - 0.3 \, \Medd$ ~ \citep{rev-2008-gal-mun-hart-psal-chak}.  The models further predict decreasing burst recurrence time with increasing accretion rate, until the point of stabilization.  Many sources, however,  after a first regime in agreement with this prediction (hereafter R1), exhibit a second regime (R2) where the recurrence time turns over and {\it increases} with increasing accretion rate, before burst quenching \citep{art-2003-corne-etal}.  Resolving this disagreement is crucial because the bursts are a critical source of information about the interior of neutron stars \citep{rev-2012-watts} and the nuclear reactions taking place on the surface, but this information can only be extracted if the burning conditions are understood.

Despite many efforts,  a firm solution to this discrepancy has not yet been found.  Most attempts involved modifying the physics already included in the models, including the uncertainties in the nuclear reaction rates \citep{art-2006-coop-nara-a,art-2014-keek-cyb-heger,art-2016-cyb-etal,rev-2018-meis-etal}. However, these modifications in fact increased the disagreement with observations.  Other approaches considered modifications to the temperature of the burning layers, since this property regulates the nuclear burning. Even though following a long series of bursts had some effect on the recurrence time \citep{art-2004-woos-etal,art-2018-johnst-etal}, stabilization at low accretion rates could only be achieved adding an extra, unidentified source of heat\footnote{Slightly counterintuitively, a very hot neutron star will not ignite explosively, because most of the fuel will be consumed by steady burning before the nuclear runaway.}.  Additional heat sources could arise from magnetohydrodynamical instabilities \citep{art-1999-ino-suny,art-2010-inoga-suny,art-2007-piro-bild,art-2009-kee-lang-zand} or a form of what is called shallow heating, an element used to fit the light curves of cooling neutron stars \citep{art-2009-brow-cumm,rev-2017-wij-dege-page}.  

Another approach considered how the accreted material is distributed over the star's surface, changing $\md$.  It has been speculated that in the second regime R2, the area covered by the accreted flow could increase with increasing accretion rate, such that $\md$ actually decreases, thus increasing the recurrence time \citep{rev-2003-2006-stro-bild}.  Rotation could also play a role, because of the high ($\sim200$--600~Hz) spins of many sources \citep{art-2007-coop-nara-a,art-2017-cavecchi-etal,art-2018-gal-etal,art-2020-cavecchi-etal}.  Coriolis force could confine most of the fluid where it is accreted near the equator, thus increasing the local accretion rate  with respect to the average over the star.  This equatorial band would determine the burst recurrence time, but stabilize very early.  After that, corresponding to the turnover into regime R2, the equator would burn stably, and ignition would be triggered at higher latitudes.  This scenario is  particularly appealing because there are indications that during the second regime, a part of the star is burning stably \citep{art-2015-lyu-etal,art-2020-cavecchi-etal}.  It has also been noted that the burst energy efficiency per accreted mass worsens significantly in the second regime \citep{art-1988-par-pen-lew,art-2020-cavecchi-etal}, possibly because the burning fuel is ``polluted'' by ashes of previous bursts and the stable equatorial burning \citep{art-2007-piro-bild,art-2020-cavecchi-etal}.

To test these ideas, we developed a framework in which both the local accretion rate and the fuel composition can be independently modified.  We calculated new burst models with the code \me{} \citep{mesa1}.  The local accretion rates are $\mdt$ from $\sim 0.04 - 1.5\, \medd$ and we use nonstandard mixtures ranging from solar to pure ashes (parametrizing them with a purity factor, $w$, from $1$ to $0$).  Composition is a key parameter to capture the effects of pollution.  We then fit the burst recurrence time and energy efficiency for the sources KS 1731$-$260, Aql X-1, EXO 0748$-$676, 4U 1636$-$536 and 4U 1608$-$522 because we have ample data from the \minb{}  catalog \citep[data release 1, version 8;][]{rev-2020-gall-etal,minbarvers8} resolving both regime R1 and R2.

In \sect{sec:mesa} we describe our numerical methods, and in \sects{sec:data} and \ref{sec:res} we describe the fitting procedure and its results. We propose a new solution to the problem in \sect{sec:sol}. Finally, in \sect{sec:conc} we draw our conclusions.

\section{MESA simulation setup}
\label{sec:mesa}

We set out to test whether a combination of different local accretion rates and compositions can explain the drop in burst recurrence time that is observed.  As a first step,  we calculated a  database of bursts over a grid of accretion rates and compositions with the code \me{} \citep{mesa1,mesa2,mesa3,mesa4,mesa5,mesa6}.  To mimic the effect of mixing ashes into the burning layers, we instead specify the composition of the accreted material, including ash species. We also set the global mass accretion rate $\MD$, which is then averaged across the surface, effectively setting $\md$. This is because \me{} considers the accretion to be homogeneous.

Before proceeding, we also want to clarify what we mean by different accreted compositions. First of all, we consider a fiducial accreted composition of solar type: $\mafra{H,solar} = 0.7$, $\mafra{He,solar} = 0.28$, $\mafra{\ele{C}{12},solar} = 0.006$, $\mafra{\ele{N}{14},solar} = 0.01$ and  $\mafra{\ele{O}{16},solar} = 0.004$, where $\mafra{\textit{i}}$ is the mass fraction of species $i$. This material could be mixed with ashes from the previous bursts from the deeper layers due to stronger rotational mixing at higher latitudes \citep{art-2007-piro-bild} or the same fluid dynamics that allows the fluid to move toward the poles. Also, in the case of a high local accretion rate at the equator, some fraction of the accreted material will burn stably before being engulfed by the burst flame ignited at higher latitudes. The specific details of the composition resulting from all these processes can only be calculated numerically with multi-D simulations.  Here, however, we mimic the effect of ``degrading'' the solar composition mentioned above by taking a weighted average between it and a fiducial ash composition.  Effectively, our accreted composition is given by
\begin{equation}
  \mafra{i} = w \mafra{\textit{i}, \rm{solar}} + (1 - w) \mafra{\textit{i}, \rm{ashes}}
  \;,
  \label{equ:comp}
\end{equation}
where $w$ is the purity factor and $w=1$ corresponds to pure solar composition and $w=0$ to pure ashes composition.

Since our goal is to mimic an impediment in the burning, in contrast to possible cases where the ashes might also contribute some energy to the burning, in this work we only mix ashes composed of pure $\ele{Kr}{80}$, the end point of our nuclear network, and hence inert to most nuclear reactions except weak decays. Therefore, $\mafra{\textit{i}, \rm{ashes}} \equiv \mafra{\ele{Kr}{80}} (= 1)$ and the input compositions resulting from \equ{equ:comp} only use H, He, $\ele{C}{12}$, $\ele{N}{14}$, $\ele{O}{16}$, and $\ele{Kr}{80}$. We explore values of $w$ between 0.1 and 1, at intervals of 0.15. We do not sample $w = 0$ since this would result in no bursts at all.

These and other (numerical) parameters are set in a so-called \texttt{inlist} file. Apart from the accretion rate and composition, we use the same \texttt{inlist} as described in  \citep{art-arx-2024-navaca-etalb}. In particular, we use the network dubbed \appr{,} which includes nuclides up to $A=80$ and covers the most relevant reaction chains for H and He burning during X-ray bursts, such as the hot CNO cycle (which reaches equilibrium composition; see, e.g., \citealt{1983PASJ...35..491H}, close to the surface, above the explosion depth), rp-process, 3$\alpha$ and $\alpha$ captures \citep{art-arx-2024-navaca-etalb}. The end point of this network is, as already said, $\ele{Kr}{80}$.

For the opacity, we use a customized routine from Bill Wolf's website\footnote{\url{https://billwolf.space/projects/leiden_2019/}}. For the radiative part, we use a combination of analytic fits from  \citep{art-1999-schatz-bild-cumm}, electron scattering from  \citep{art-1983-pac-b}, and the corrective factor from  \citep{art-2001-potek-yakov}. For the conductive part, we employ the tables provided by \texttt{MESA}. General relativity corrections are implemented by multiplying the gravitational constant $G$ by the corrective factor $(1 - 2GM/c^{2}r)^{-1}(1 + 4\pi r^{3}P/mc^{2})(1 + P/\rho c^{2})$.

\me{} needs an initial stellar model representing the outer layers of the neutron star, where accretion will take place, and a luminosity value at the lower boundary that will be constant during the simulation. We construct such models using the code described in  \citep{art-arx-2024-navaca-etala}. We start the integration at a density of $\rho = 10^4$ g cm$^{-3}$ until a lower boundary at $\rho = 10^9$ g cm$^{-3}$. These models are composed of pure $\ele{Kr}{80}$. Given a value of effective temperature $\teff$ at the surface, these models will also provide a value for the base luminosity, $\Lb$, at the lower boundary.

As for the base luminosity for \me{,} we make it proportional to the mass accretion rate of choice, according to $75\,$keV per accreted nucleon. This numerical factor is lower than typically used.  In general, there will be reactions in the deeper layers of the neutron star crust, the so-called deep crustal heating, which liberate approximately $0.6 - 2.7$ MeV per accreted nucleon, a fraction of which will go into the core and be lost via neutrinos \citep{art-1990-haen-zdun,art-2008-haen-zdun,art-2008-gupta-etal.pdf,art-2018-fort-etal,art-2018-lau-etal,art-2021-gusa-chugu,art-2022-shche-etal,art-2023-pote-etal,art-arx-2025-navaca-etal}.  On top of this, there would be the shallow heating, which can generate up to 2 or 3 MeV per accreted nucleon in the outer crust \citep[see, e.g.,][and references therein]{art-arx-2025-navaca-etal}.  How much of the total heat will go into the envelope and how much will go into the core depends on the full structure of the star and its accretion history. The base flux for the bursts could even be negative in the case of a cold star, or it could be extremely large in the presence of a strong shallow heating, covering a possible range of $\sim -0.7 - 2.7$ MeV per accreted nucleon \citep{art-arx-2025-navaca-etal}. Given this large uncertainty, we do \emph{not} use high luminosities $\Lb$ because our first goal is to see if we can reproduce the observed early quenching of the bursts mainly with compositional effects related to accretion physics.

One important point for the numerical stability of our simulations was that the temperature structure of the initial models should be consistent with the mass accretion rate and the base luminosity. In the code of  \citet{art-arx-2024-navaca-etala}, the base luminosity is a product of the integration and can only be controlled by the choice of $\teff$ once the composition and the range of density are chosen. Thus, we initially prepared a set of models with various $\teff$, recorded their $\Lb$, and subsequently interpolated among these values to obtain the $\teff$ corresponding to our $\Lb$ values of choice (set by $\MD$). We calculated the stationary models corresponding to these $\teff$ and used them as initial conditions. \me{} initially evolved these models to equilibrium, and then we started accretion and the burst simulations.

\subsection{Burst model database}
\label{sec:burstdb}

\begin{table}
  \centering
\begin{tabular}{ll|ccr}
$\MD$     & $w$ & $t_{\rm{rec}}$ & $\alpha$ & $N_{\rm{b}}$\\{}
($\Medd$) &     & (hr)           &          &             \\
\hline
$1.54          $ & 1.00 & $1.94 \pm 0.08\, 10^{-1}$ & $3.39 \pm 0.74\, 10^{1}$ & 10\\
$1.54          $ & 0.85 & $1.65 \pm 0.12\, 10^{-1}$ & $4.09 \pm 0.76\, 10^{1}$ & 10\\
$1.54          $ & 0.70 & $1.69 \pm 0.13\, 10^{-1}$ & $5.23 \pm 0.76\, 10^{1}$ & 16\\
$1.54          $ & 0.55 & $1.28 \pm 0.29\, 10^{-1}$ & $7.80 \pm 2.63\, 10^{1}$ & 72\\
$1.54          $ & 0.40 & $7.22 \pm 2.05\, 10^{-2}$ & $9.28 \pm 6.65\, 10^{1}$ & 16\\
$1.54          $ & 0.25 & $7.74 \pm 2.82\, 10^{-2}$ & $2.05 \pm 1.52\, 10^{2}$ & 45\\
$1.54          $ & 0.10 &  --                       & --                       & --\\
$6.17\, 10^{-1}$ & 1.00 & $5.45 \pm 0.28\, 10^{-1}$ & $3.64 \pm 0.52\, 10^{1}$ &  9\\
$6.17\, 10^{-1}$ & 0.85 & $3.82 \pm 0.23\, 10^{-1}$ & $4.34 \pm 0.98\, 10^{1}$ &  9\\
$6.17\, 10^{-1}$ & 0.70 & $3.23 \pm 0.20\, 10^{-1}$ & $5.24 \pm 1.14\, 10^{1}$ &  8\\
$6.17\, 10^{-1}$ & 0.55 & $3.04 \pm 0.08\, 10^{-1}$ & $6.84 \pm 0.87\, 10^{1}$ & 19\\
$6.17\, 10^{-1}$ & 0.40 & $3.06 \pm 0.12\, 10^{-1}$ & $9.62 \pm 1.57\, 10^{1}$ & 11\\
$6.17\, 10^{-1}$ & 0.25 & $3.49 \pm 0.16\, 10^{-1}$ & $1.61 \pm 0.23\, 10^{2}$ &  9\\
$6.17\, 10^{-1}$ & 0.10 & $4.17 \pm 0.22\, 10^{-1}$ & $4.80 \pm 0.66\, 10^{2}$ & 16\\
$2.47\, 10^{-1}$ & 1.00 & $1.44 \pm 0.08          $ & $4.11 \pm 0.31\, 10^{1}$ & 10\\
$2.47\, 10^{-1}$ & 0.85 & $9.96 \pm 0.59\, 10^{-1}$ & $4.80 \pm 0.58\, 10^{1}$ &  8\\
$2.47\, 10^{-1}$ & 0.70 & $9.01 \pm 0.58\, 10^{-1}$ & $5.96 \pm 0.75\, 10^{1}$ &  8\\
$2.47\, 10^{-1}$ & 0.55 & $9.23 \pm 0.78\, 10^{-1}$ & $7.58 \pm 1.26\, 10^{1}$ & 11\\
$2.47\, 10^{-1}$ & 0.40 & $9.54 \pm 0.52\, 10^{-1}$ & $1.12 \pm 0.14\, 10^{2}$ & 24\\
$2.47\, 10^{-1}$ & 0.25 & $1.10 \pm 0.04          $ & $1.90 \pm 0.14\, 10^{2}$ &  9\\
$2.47\, 10^{-1}$ & 0.10 & $1.58 \pm 0.28          $ & $6.55 \pm 1.43\, 10^{2}$ & 12\\
$9.87\, 10^{-2}$ & 1.00 & $3.69 \pm 0.05          $ & $4.93 \pm 0.25\, 10^{1}$ & 10\\
$9.87\, 10^{-2}$ & 0.85 & $2.70 \pm 0.13          $ & $5.65 \pm 0.40\, 10^{1}$ &  8\\
$9.87\, 10^{-2}$ & 0.70 & $2.42 \pm 0.08          $ & $6.82 \pm 0.87\, 10^{1}$ &  6\\
$9.87\, 10^{-2}$ & 0.55 & $2.19 \pm 0.19          $ & $9.28 \pm 1.46\, 10^{1}$ &  9\\
$9.87\, 10^{-2}$ & 0.40 & $2.80 \pm 0.26          $ & $1.38 \pm 0.28\, 10^{2}$ &  7\\
$9.87\, 10^{-2}$ & 0.25 & $3.02 \pm 0.09          $ & $2.83 \pm 0.53\, 10^{2}$ &  7\\
$9.87\, 10^{-2}$ & 0.10 & $5.16 \pm 0.31          $ & $9.56 \pm 1.29\, 10^{2}$ &  9\\
$3.95\, 10^{-2}$ & 1.00 & $8.71 \pm 0.75          $ & $6.29 \pm 2.10\, 10^{1}$ &  7\\
$3.95\, 10^{-2}$ & 0.85 & $6.89 \pm 0.29          $ & $7.95 \pm 0.75\, 10^{1}$ &  6\\
$3.95\, 10^{-2}$ & 0.70 & $5.97 \pm 0.60          $ & $1.06 \pm 0.37\, 10^{2}$ &  7\\
$3.95\, 10^{-2}$ & 0.55 & $6.04 \pm 0.45          $ & $1.33 \pm 0.33\, 10^{2}$ & 11\\
$3.95\, 10^{-2}$ & 0.40 & $8.30 \pm 0.78          $ & $1.67 \pm 0.67\, 10^{2}$ &  5\\
$3.95\, 10^{-2}$ & 0.25 & $1.81 \pm 1.42\, 10^{ 1}$ & $8.10 \pm 7.15\, 10^{2}$ &  5\\
$3.95\, 10^{-2}$ & 0.10 &  --                       & --                       & --\\
\end{tabular}
  \caption{Simulated burst database values. Values for accretion rate, $\mdo$ (1.54, $6.17\, 10^{-1}$, $2.47\, 10^{-1}$, $9.87\, 10^{-2}$, $3.95\, 10^{-2}$ $\medd$), the mass fraction of ashes, $w$ (1, 0.85, 0.70, 0.55, 0.40, 0.25, 0.1), the burst recurrence time, $t_{\rm{rec}}$ and $\alpha$ with their respective 1 $\sigma$ errors and the number of bursts used.  $\alpha$ is calculated according to \equ{equ:alpha}. All values are calculated at the star surface without any redshift correction. When data are not available, the simulations displayed an oscillatory behavior reminiscent of mHz QPOs. $\medd$ here is intended for a composition with $\mafra{H} = 0.7$, and a star with $M_* = 1.4 M_\odot$ and $R_* = 11.4\,$Km $\Medd \sim 2 \, 10^{-8} \rm{M}_\odot / \rm{yr}$.}
  \label{tab:tibs}
\end{table}

For our burst database, the values of $\MD$ were logarithmically equispaced, while the ash fraction values $w$ were linearly spaced. Their values and the resulting burst parameters are summarized in \tab{tab:tibs}.  In particular, we report the burst recurrence time and the nuclear energy release per unit accreted mass.  This is quantified as customary by the $\alpha$ parameter, which compares the nuclear energy release in a burst to the gravitational energy release of accretion between bursts,
\begin{equation}
  \alpha = \frac{Q_{\rm{grav}}}{Q_{\rm{nuc}}}
  \label{equ:alpha}
  \;,
\end{equation}
where $Q_{\rm{grav}} \sim G M_{\rm{NS}} / R_{\rm{NS}}$, and $Q_{\rm{nuc}}$ depends on the composition of the fuel layer at ignition \citep{art-2019-good-etal}, as well as on the fraction of the accreted material participating in the burts --- we recall, much of the material may burn in steady state at the equator. Observationally, this is measured by integrating the fluence of the persistent emission compared to the fluence of the bursts, eliminating some uncertainties such as the distance to the source.  Less efficient bursts will have higher $\alpha$.

\figudb{}

In \fig{fig:db}, left, we show the evolution of the burst recurrence time as a function of the weight $w$ for each of our mass accretion rates $\MD$. Analogously, in \fig{fig:db}, right, we show the $\alpha$ parameter as a function of $\MD$ and $w$.
There is a trend that immediately strikes the eye. Contrary to intuition, the burst recurrence time initially \emph{decreases} when we begin to mix ashes in the composition. This is counterintuitive because it means that adding ashes actually improves the burning to the point that the bursts ignite earlier than in the pure solar composition case. After the ashes constitute a high enough fraction of the accreted material, this trend inverts, and the recurrence time begins to increase, leading to less frequent bursts.

We are convinced that this trend is real and not a numerical artifact. Indeed, although not mentioned in previous works \citep{art-2000-cumm-bild,art-2016-lampe-heg-gal,art-2019-good-etal,art-2020-johnst-etal}, this trend is visible there as well (for example, in the simulations of  \citealt{art-2019-good-etal}). It was not mentioned because typically the changes in composition were limited to H, He, and CNO elements, which have a smaller impact on the opacity. We ran a series of tests changing the network and the kind of ashes we use, and confirmed this trend is real and not a numerical issue. These interesting effects of the ashes composition on burst properties should be explored further.

As a curiosity, we note that the opacity can stabilize the bursts \citep{art-arx-2024-navaca-etalb}.  If the energy transport is somewhat impeded in the layers \emph{above} the burst ignition, the burning turns stable at accretion rates lower than the Eddington rate.  We think that there is a similar effect in our simulations due to the composition. Since we add $\ele{Kr}{80}$ to the burning mixture, its high $Z$ number increases the opacity \citep{art-1999-schatz-bild-cumm,art-2001-cumm-bild}.  Then, what we see is the competition of two effects. On the one hand, the increased opacity helps the explosion by decreasing the time to ignition. On the other hand, diminishing the amount of H, He, and CNO elements, we are also reducing the steady burning energy release between bursts.  The competition between the two effects produces the initial decrease, followed by an increase in the burst recurrence time as we progressively pollute the fluid composition. Note, however, that this effect is not comparable to the trend in the data, and we are by no means implying this can explain the observations.

\section{Data analysis}
\label{sec:data}

In order to gain insight into the local conditions that determine the burst properties as the mass accretion rate evolves, we can reverse engineer the data, finding the pair $(\mdt, w)$ which returns values compatible with the pair $(\tr, \alpha)$ of each data point, where $\tr$ is the recurrence time.  We apply our burst database to the sources KS 1731-260, Aql X-1, EXO 0748-676, 4U 1636-536 and 4U 1608-522 extracting their properties from the \minb{} database \citep{rev-2020-gall-etal}
and take the same approach of \citet{art-2018-gal-etal} to estimate the $(\tr, \alpha)$--values as a function of the accretion rate, as follows.

From \minb{} we collect all the available information about the observations for a given source. We then divide these data into bins that are roughly equispaced in mass accretion rate (or rather a proxy for it, $\gamma$, which is the ratio of the persistent flux to the average Eddington flux for each source; see \citealt{rev-2020-gall-etal} for more details). The exact bin ranges depend on the condition to have enough bursts in each bin (50 in our case, apart from Aql X-1 and 4U 1608$-$522 for which we require 25 data points per bin). We also collect the whole dataset of the same source in the same bins, independently of whether the observations contain bursts or not, and derive $\MD$ from the average of the persistent emission. 

Finally, for each bin, we calculate the average value of the burst rate $\rb$ (which we convert into the recurrence time) and $\alpha$. Notably, we correct the observational data to take into account the different anisotropy effects on the burst and persistent emission according to inclination ranges from the literature, combined with the modeling of \cite{he16}, so that the values reported are directly comparable to our simulations. All the data are summarized in \tab{tab:data}.

As for turning our discrete grid of theoretical values of $\tr$ and $\alpha$ into a continuous function of $\mdt$ and $w$, we set up a bilinear interpolator of our database results. Then, for each bin \emph{independently}, we first find the best model parameters $\mdt$ and $w$ which minimize the residuals between the interpolator results and the data (through least squares minimization). The function we minimize is proportional to the normalized distance to the observed values:
\begin{equation}
\frac{(t_{\rm{rec, model}} - t_{\rm{rec, obs}})^2}{2 \sigma_{t_{\rm{rec, obs}}}^2} + 
\frac{(\alpha_{\rm{model}} - \alpha_{\rm{obs}})^2}{2 \sigma_{\alpha_{\rm{obs}}}^2}
\label{equ:chi2}
\end{equation}
The observational errors were always larger than the theoretical ones in the regions of best fit. The errors on $\mdt$ and $w$ are calculated using the Markov Chain Monte Carlo code \texttt{emcee} \citep{art-2013-foremack-etal}.  We assume flat priors between 0 and $\mdstab$ for $\mdt$ and between 0 and 1 for $w$.  The log-likelihood is given by the negative value of \equ{equ:chi2}, where we ignore unimportant constants. We seed 100 walkers randomly distributed according to a normal distribution centered on the best fit values, with a standard deviation based on the error of the least square minimization, and let them proceed for 100,000 steps. We tried longer chains, but it was not necessary.

\section{Results}
\label{sec:res}

The results of the fit are shown in \fig{fig:res} for the sources KS 1731-260 and 4U 1636-536 and \fig{fig:resM} for all the sources. Our results defy all previously mentioned models, but they allow us to find a new solution.  We plot composition, through $w$, and local accretion rate $\mdt$.  To infer the conditions on the star that control the burning, it is useful to also plot the ratio between the (latitude-dependent) accretion rate $\mdt$ and the average local rate $\mdo$:
\begin{equation}
  f(\theta) = \frac{\mdt(\theta)}{\mdo} =
              \frac{\MD_{\rm{local}}(\theta)}{\Alo} /
              \frac{\MD_{\rm{global}}}{4\pi R_*^2} =
              \mfr(\theta) \frac{4\pi R_*^2}{\Alo}
\end{equation}
where $\mfr(\theta) = \frac{\MD_{\rm{local}}(\theta)}{\,\MD_{\rm{global}}}$.
Two things contribute to $f(\theta)$ in a certain region around latitude $\theta$: how much of the accreted matter arrives there, $\mfr$, and the area of the region, $\Alo$.  Note that $f$ can be larger than one due to a small $\Alo$ and $\mfr \sim 1$, or it can be much smaller, if $\mfr \ll 1$.

\resufigB{}

First, we focus on the decreasing burst recurrence time regime, R1.  In the top two rows of \fig{fig:res}, before the turnover, we can see that the composition is constant (and often practically solar, $w \sim 1$; see also \fig{fig:resM}).  $\mdt$ increases, but it tracks closely $\mdo$ and their ratio $f$ is initially roughly constant as well, \emph{with signs of a slight increase}.  Since this ratio is greater than one, we conclude that most of the accreted material is distributed over a limited fraction of the star's surface, which most likely corresponds to low latitudes (see next section).  However, the turnover takes place before ever reaching $\mdstab$, so we can exclude that the transition is only due to the rotation and surface conditions as proposed earlier \citep{art-2020-cavecchi-etal}.

At accretion rates above the turnover, R2, both $f$ and $w$ decrease  dramatically.  Allowing all the fluid to suddenly spread over a wider area \citep{rev-2003-2006-stro-bild} would explain why $\mdt$ decreases during the second regime and therefore never reaches $\mdstab$.  However, the ratio $f$ should not fall below $1$, because that limit corresponds to the area of accretion coinciding with the entire star surface.  That is because if we do not reach stabilization, we always have the whole accreted matter at disposal for burning, and $\mfr$ should always be around 1, as should $w$.  This is in stark contradiction with our findings.  We tested whether we could reproduce the data allowing only the composition to change in the second regime, but we could not find a good fit (composition must change because of the lower energy efficiency in R2).  This demonstrates that not only the composition worsens, but the location that leads to the burst ignition changes and receives a decreasing fraction of the accreted material.  As discussed below, these $f$ and $w$ would correspond to high latitudes.

%%%%%%%%%%%%%%%%%%%%%%%%%%%%%%%

It is important to stress that these trends are not a result of parametrization: as we mentioned, we fit each bin independently from the others, so the trends are only dictated by the data, and all sources follow the same trend. We report the values of the approximate turnover and the weighted averages of $f$ and $w$ before the drop in \tab{tab:res}.

We note that for the sources Aql X-1 and EXO 0748-676 the initial weight $w$ is not 1, but it is still constant and subsequently drops. This is easily explainable with an accretion composition which is different from solar, in particular, probably poorer in hidrogen. This is similar to what has been found for IGR J17498-2921 and SAX J1808.4-3658 \citep{art-2024-gall-etal}. Aql X-1 and EXO 0748-676 follow nonetheless the same trend as the others (see \figs{fig:resM} and \ref{fig:resS}).

For the source EXO 0748-676, the value of $f$ before the turnover is $< 1$.  This value is clearly an outlier with respect to the others, while the values of the local $\mdt$ are not. The difference is only due to the values of $\mdo$ from \minb{.} EXO 0748-676 is the only dipper in our sample, and it is possible that \minb{} applies corrections that overestimate $\mdo$.

To be sure, we also ran some tests using a database made with the \kep{} code. Even though the numerical values of $f$ and $w$ and their averages before the turnover may vary slightly, all the fits qualitatively agree on the same trend\footnote{The database of \kep{} had some troubles reproducing the data of Aql X-1 and EXO 0748-686, probably due to the fact that the composition of these sources needed to be substantially different from solar at all mass accretion rates, a possibility not included in the database.}.

Finally, one could argue that $\mdstab$ is reached, and the rate does actually reach the maximum predicted, but we do not see this in the data because this happens in a very short range of $\MD$ and is washed out by the averages in the bins.  We have checked with different binning conditions, and we did find possibly a higher peak burst rate, but never close enough to the maximum rate expected from simulations, and with larger error bars.
%%%%%%%%%%%%%%%%%%%%%%%%%%%

\begin{table*}
  \centering
  \begin{tabular}{l|llllll}
    Source & $\ave{f}$ & $\sigma_{\ave{f}}$ & $\ave{w}$ & $\sigma_{\ave{w}}$ & $\mdto/\medd$ & $S_{z, \, \rm{to}}$\\
    \hline
    KS 1731     & 1.799 (1.746) & 0.206 (0.175) & 0.990 & 0.021 &  8.968$\, 10^{-2}$ & 1.298 \\
    Aql X-1 & 1.730 & 0.620 & 0.288 & 0.030 &  6.689$\, 10^{-2}$ & 1.763 \\
    EXO 0748     & 0.703 & 0.068 & 0.391 & 0.044 &  1.558$\, 10^{-1}$ & 1.226 \\
    4U 1636     & 3.366 (2.155) & 0.431 (0.205) & 1.000 (0.983) & 0.023 (0.022) &  5.005$\, 10^{-2}$ & 1.385 \\
    4U 1608     & 1.691 (1.601) & 0.334 (0.298) & 0.949 (0.938) & 0.033 (0.031) &  5.151$\, 10^{-2}$ & 1.532 \\
  \end{tabular}
  \caption{Average values and their errors for $f$ and $w$ before the turnover. The last columns indicate the approximate position of the turnover in observed mass accretion rate $\mdto$ and $S_{z, \, \rm{to}}$. They correspond to the first bin after the turn over. Averages are done only for bins with $\mdo < \mdto$ (numbers in parentheses are obtained with ordering based on $S_z$ when those are different, but they are always compatible within $3 \sigma$).}
  \label{tab:res}
\end{table*}

\section{A unified solution: the accretion states}
\label{sec:sol}

\figscheme{}

We find that the observations can be explained as arising from the change in the disk structure around the turnover. It is the disk that determines which regions can ignite and which ones burn steadily, polluting the others.
One important observational fact informing our proposed model is that when the accretion rate increases, the persistent spectra of the burst sources transition from being predominantly hard (island state) to soft \citep[banana state;][]{art-1989-hasi-klis}. These spectral states are thought to correspond to a change in the disk structure near the surface of the star. For this reason, we plot our results as a function of the parameter $S_z$, which tracks the spectral state, in Figure \ref{fig:res}, bottom rows, and in \fig{fig:resS}. 
For selected sources, \minb{} lists this parameter \citep{rev-2020-gall-etal}, which tracks the position of the source in the X-ray ``color-color'' diagram. For each observation,  a ``hard'' and ``soft'' color is calculated, usually as the ratio of instrumental counts measured in two bands, each in higher (hard) or lower (soft) energy ranges. $S_z$ is the parametrised position within the track drawn out by a source over many observations, imposing that $S_z = 1$ at the end of the hard and $S_z = 2$ at the beginning of the soft state. To each bin, we assign its average value.
It is believed that $S_z$, on top of following the state, tracks $\MD$ better than the persistent emission \citep{art-2007-lin-etal}, so this may be an even better ordering of the data points.
The same trend is visible in these plots. Actually, the evolution of $f$ before the turnover is perhaps more evident.  Very importantly, the turnover takes place in a range of $S_z \simeq 1.2 - 1.7$ for all the sources we considered (\tab{tab:res}). These values correspond to a position midway between the hard state (1) and the soft state (2), while the first regime corresponds to the hard state.

One key difference between the accretion disks in these two states is the presence near the star of a corona or hot inner flow (hot Comptonizing gas) in the hard state.  This flow is relatively puffed with respect to a normal thin disk and comparable to a sizable fraction of the star radius \citep{rev-2007-done-etal}.  When the corona weakens toward the soft state, the inner disk thins.  It develops a boundary layer where it connects to the star and an equatorially concentrated spreading layer over its  surface \citep{art-1999-ino-suny,art-2010-inoga-suny}.

Although the inner disk structure is not yet clearly established, the configurations described above are supported by a growing amount of evidence \citep{rev-2007-done-etal}.  For example, the spectral evolution of Aql X-1 and 4U 1608$-$522 \citep{art-2007-lin-etal} showed that the temperature of a blackbody component with roughly constant scale height was increasing and tracking the mass accretion rate onto the neutron stars. The authors identified this component with a boundary layer of growing temperature, while the comptonizing medium contribution weakens significantly toward the soft state.  The presence of the spreading layer in the soft state was identified in the spectra of 4U 1608$-$522 \citep{art-2017-kaja-etal}.
A growing boundary and spreading layers also regulate the matter flow during the state evolution of the source 4U 1820-30 \citep{art-2023-mari-etal}.  Here the boundary layer in the very early soft state anticorrelates with the presence and intensity of a jet, probably due to impoverishment of the corona.

If we consider this state transition and the associated inner disk change, combined with our results, we reach an unexpected, but actually natural, explanation for the burst behavior (see \fig{fig:scheme}).  Initially, \textit{a}, the accretion is more uniform due to the hot inner accretion flow \citep{rev-2007-done-etal,art-2014-kaja-etal} and it covers a large, but still $<1$, fraction of the star surface.  When the accretion rate reaches the threshold for state transition, \textit{b}, the disk begins to switch to a spreading layer configuration, and the accretion flow becomes more concentrated around the equator.  For this transition to happen, the height of the accretion flow adjacent to the star must have a smaller opening angle than the spreading layer would have at the same luminosity \citep{art-2010-inoga-suny}.  Thus, for the turnover to happen, the corona needs to diminish, and the fluid should actually cover a \emph{smaller} area.  Finally, \textit{c}, the spreading layer is fully formed.

This explanation is consistent with the constant composition $w$ and constant, or slightly evolving, $f$ at accretion rates below the turnover (R1, \textit{a}).  A larger initial area means a smaller $f$, when this area slowly shrinks, $f$ slightly increases.  The burning composition is the same as that of the donor, roughly solar.  At accretion rates above the turnover, R2, this scenario explains the rise in burst recurrence time and drop in energy efficiency, or, equivalently, the drop in $f$ and purity $w$, by the effects of the spreading layer distribution, \textit{c}.  In particular, the spreading layer has higher temperature near the equator, and a local accretion rate close to Eddington \citep{art-2006-sulei-pout,art-2010-inoga-suny}.  This implies stable burning that would consume most of the accreted material and also pollute the composition of the fuel left to ignite.  The burst behavior would be dictated mostly by the small fraction of fuel that reaches higher latitudes (thus reducing $f$ and $w$ at the same time).  Also note that the spreading layer reaches the pole at roughly half the Eddington global accretion rate \citep{art-2006-sulei-pout}. This is compatible with the luminosity at which the bursts disappear completely, once again in agreement with the idea that there is no more area of the star surface capable of igniting.  Finally, another factor that may affect the fraction of mass available is the launching of hot winds.  This occurs preferentially near the soft state  \citep{rev-2023-neil-dege}, even if optical and infrared winds have also been detected in the hard state \citep{art-2020-sasi-dari}.

We argue that this scenario can unify the whole spectrum of burster behaviors within a single, natural mechanism.  On one hand, without adding unknown physics, we showed that sources where the bursts are quenched at significantly lower accretion rates than normally predicted can be brought back in agreement with nuclear physics calculations.  On the other hand, the sources that follow the classical models may do so simply because they never develop a stable boundary layer, either because they never reach the necessary accretion rate\footnote{A classic example is the clocked-burster, GS 1826-238, which for many decades has been thought to be very regular and well described by 1D simulations. However, all its bursts had been detected in the hard state. Once the source switched to the soft state, its bursts also started deviating from expectations with increasing recurrence time \citep[see, e.g.,][and references therein]{2025ApJ...993L..13T}.} or because their disks are truncated.  Advanced numerical simulations have begun to show how weak and strong magnetic fields will affect the inner disk structure \citep{art-2017-parf-tchek,art-2022-das-port-watts,art-2024-parf-tchek,art-2024-das-port}, supporting this idea.   In these cases, the channelled accretion and spreading of the fluid at deeper layers \citep{art-1997-bild-brown} will change the distribution and composition of the matter on the surface, affecting the corresponding burst rate and efficiency as functions of the accretion rate.  Thus, our findings highlight the potential of the bursts as probes of another key physics: not only are they useful to gain information about nuclear physics and the interior of neutron stars, but also about the yet unclear accretion processes near their surfaces.

\subsection{Implications for other observables}

If the reason behind the turnover is mostly due to a change in the acccretion disk, we need to revise the interpretation of two more observational facts: the presence of the mHz quasiperiodic oscillations (mHz QPOs) concurrently with the bursts and the inverse relation between the turnover mass accretion rate and the spin of the neutron stars. 

\subsubsection{The implications for mHz QPOs}

The mHz QPOS were first reported in 2001 and were immediately suggested to be associated with nuclear burning \citep{art-2001-rev-etal}.  These QPOs have been identified with an oscillatory burning regime predicted to appear when the accretion conditions are close to stabilization \citep{art-1983-pac-a}, using numerical simulations with the multizone, 1D code \kep{} and the single zone code \textsc{onezone} \citep{art-2007-heg-cumm-gal-woos}.  Note also that there are hints that the mHz QPOs are emitted by a narrow region and not the whole surface, although this is debated \citep{art-2018-stroh-etal,art-2020-hsieh-chou}.

The presence of the mHz QPOs from the turnover on in the case of 4U 1636-536 \citep{art-2014-lyu-etal,art-2015-lyu-etal,art-2016-lyu-etal} can be accounted for as the result of quasi-stable burning in a zone between the stabilized region near the equator and the still unstable higher latitudes, although not because of a region confined solely by the Coriolis force as previously suggested \citep{art-2020-cavecchi-etal}.

\subsubsection{The implications for the mass accretion rate of turnover}

On the contrary, the reported inverse dependence of the turnover mass accretion rate on rotation spin \citep{art-2018-gal-etal,art-2020-cavecchi-etal} becomes now slightly problematic.  \tab{tab:res} shows the same result. The turnover mass accretion rate decreases by a factor of $\sim 2$ \footnote{Even accounting for the general relativistic corrections to the conversion efficiency between $\MD$ and luminosity due to the star rotation \citep{art-2000-sibga-suny}, this factor does not change significantly.} when the spin frequency increases from $\sim 500$ Hz to $\sim 600$ Hz.
At the moment, it is not clear what effect rotation can have on the accretion disk changes, if any.

From the point of view of the accretion flow, a faster star would have colder boundary and spreading layers \citep{art-2010-inoga-suny}. If the heat of these layers affects the burning around the equator and/or the accretion rate at which the inner hot flow disappears, this would imply that the turnover should start earlier for slower stars, contrary to what is observed.

On the other hand, from the point of view of the bursts, the accreted material onto a faster star would have less angular momentum with respect to the star, and this may arguably reduce the extent toward the poles that such material would cover, both in the case of the inner hot flow and the spreading layer.  This is consistent with the observations.

One could speculate that a factor that could slow down the accretion geometry transition for slower stars could be the fact that the equatorial radius of slower rotating stars is shorter and that possibly the slower rotation may lead to a slower depletion of the corona, compared to faster spins, by slowing (the onset of) the disk wind, even if through which coupling is difficult to say. The interplay of the processes removing the corona and concentrating the fluid near the equator may thus still push in the direction of reducing the turnover mass accretion rate for faster rotators, but more investigations, both theoretical and observational, are needed.

\section{Conclusions}
\label{sec:conc}

In this work, we built a database of Type I bursts varying the mass accretion rate and the fuel composition, mixing solar composition fuel with ashes of pure $^{80}\rm{Kr}$ (the end point of our reaction network).  The resulting burst recurrence time is smaller when a small fraction of ashes is mixed in, but eventually is much higher when a high fraction is added. This is probably because the ashes initially increase the opacity in the burning layers, keeping more heat and helping ignition, but eventually worsen the energy release rate more than they help ignition.

We fit our database to observations of five sources, trying to match the burst recurrence time and $\alpha$ so to estimate the local conditions that determine the burst properties. We find that when the burst recurrence time is increasing as a function of $\mdo$, R1, the local composition is roughly constant. The local $\mdt$ regulating the bursts follows the average $\mdo$ by a factor that is roughly constant or slightly increasing toward the turnover. When the burst recurrence time increases, R2, the composition worsens progressively and the local $\mdt$ becomes a progressively smaller fraction of $\mdo$.

Given that the turnover of the burst recurrence time takes place during the transition from the hard to the soft state, we conclude that the switch of behavior is mainly dictated by the change in the accretion disk. 
The part of the disk closer to the neutron star is a hot inner flow in the hard state, which is more homogeneous and covers a larger range of latitudes. In the soft state, it switches to a spreading layer, more concentrated near the equator and hotter, causing the changes in the burst behavior by forcing stable burning at lower latitudes. This mechanism is capable of explaining all the burst phenomenology within a single picture.

Thus, Type I bursts provide a further, independent confirmation of the same paradigm that is emerging from the studies of accretion flows around neutron stars and black holes. This opens the possibility to use such bright flashes to also probe the changing accretion--ejection configuration of compact objects across a wide range of luminosities.

\section*{Acknowledgments}

We thank B. de Marco, A. Marino, and R. Wijnands for useful discussions.
L.R.S. and Y.C. dedicate this paper to the loving memory of T. G. Garfield.
We thank B. Paxton and all \me{} developers for making the code public.
Y.C. acknowledges support from the grant RYC2021-032718-I, financed by MCIN/AEI/10.13039/501100011033 and the European Union NextGenerationEU/PRTR. M.N.C.'s work is supported by the Fonds de la Recherche Scientifique-FNRS under grant No. IISN 4.4502.19. D.P. acknowledges support from a UNAM-DGAPA grant PAPIIT-IN114424.

\section*{Data Availability}
All the data used in this paper are available at \href{https://bridges.monash.edu/collections/Multi-INstrument_Burst_ARchive_MINBAR_data_release_1/4858818}{\minb{} data release webpage} \citep{minbarvers8}.
The code \me{} is available \href{https://docs.mesastar.org/en/latest/installation.html#installing-mesa}{here}. The initial models to run MESA are available under an open-source Creative Commons Attribution license \dataset[here]{\doi{10.5281/zenodo.17652518}}.
The code used to calculate the initial stationary models will be made available upon reasonable request.

\bibliographystyle{aasjournal}
\bibliography{ms.bib}{}

\appendix
\restartappendixnumbering

\section{The data}

Here, we report the data from the catalogue \minb{.}

\begin{table}
\centering
\begin{tabular}{lllllll}
\hline
Source & $\dot{m}_{\rm{obs}, \, \rm{i}} / \dot{m}_{\rm{Edd}}$ & $\dot{m}_{\rm{obs}, \, \rm{f}} / \dot{m}_{\rm{Edd}}$ & $\overline{\dot{m}_{\rm{obs}} / \dot{m}_{\rm{Edd}}}$ & $\overline{S_{z}}$ 
& $t_{\rm{rec, obs}} \, [\rm{hr}]$
& $\alpha_{\rm{obs}}$ \\
\hline
KS 1731     & $1.16\, 10^{-2}$ & $5.61\, 10^{-2}$ & $3.65\, 10^{-2}$ & $0.89$ & $5.41         $ $\pm$ $7.66\, 10^{-1}$ & $6.81\, 10^{1}$ $\pm$ $1.12\, 10^{1}$ \\
            & $5.61\, 10^{-2}$ & $6.86\, 10^{-2}$ & $6.20\, 10^{-2}$ & $1.23$ & $2.41         $ $\pm$ $3.41\, 10^{-1}$ & $3.78\, 10^{1}$ $\pm$ $6.27         $ \\
            & $6.86\, 10^{-2}$ & $8.24\, 10^{-2}$ & $7.62\, 10^{-2}$ & $1.15$ & $2.96         $ $\pm$ $4.38\, 10^{-1}$ & $4.56\, 10^{1}$ $\pm$ $7.75         $ \\
            & $8.24\, 10^{-2}$ & $8.89\, 10^{-2}$ & $8.47\, 10^{-2}$ & $1.15$ & $2.04         $ $\pm$ $2.87\, 10^{-1}$ & $3.70\, 10^{1}$ $\pm$ $6.04         $ \\
            & $8.89\, 10^{-2}$ & $1.04\, 10^{-1}$ & $9.47\, 10^{-2}$ & $1.11$ & $2.14         $ $\pm$ $3.03\, 10^{-1}$ & $4.35\, 10^{1}$ $\pm$ $7.17         $ \\
            & $1.04\, 10^{-1}$ & $1.33\, 10^{-1}$ & $1.13\, 10^{-1}$ & $1.37$ & $2.74         $ $\pm$ $3.84\, 10^{-1}$ & $5.24\, 10^{1}$ $\pm$ $8.53         $ \\
            & $1.33\, 10^{-1}$ & $2.20\, 10^{-1}$ & $1.82\, 10^{-1}$ & $1.97$ & $7.14         $ $\pm$ $9.90\, 10^{-1}$ & $3.90\, 10^{2}$ $\pm$ $6.35\, 10^{1}$ \\
            & $2.20\, 10^{-1}$ & $1.69          $ & $2.63\, 10^{-1}$ & $2.05$ & $1.12\, 10^{1}$ $\pm$ $2.93          $ & $1.10\, 10^{3}$ $\pm$ $3.05\, 10^{2}$ \\
\hline
Aql X-1     & $5.20\, 10^{-3}$ & $4.21\, 10^{-2}$ & $2.01\, 10^{-2}$ & $0.80$ & $1.34\, 10^{1}$ $\pm$ $2.70          $ & $6.28\, 10^{2}$ $\pm$ $1.32\, 10^{2}$ \\
            & $4.21\, 10^{-2}$ & $6.27\, 10^{-2}$ & $5.19\, 10^{-2}$ & $1.53$ & $3.26         $ $\pm$ $6.57\, 10^{-1}$ & $3.00\, 10^{2}$ $\pm$ $6.35\, 10^{1}$ \\
            & $6.27\, 10^{-2}$ & $1.19\, 10^{-1}$ & $8.19\, 10^{-2}$ & $1.99$ & $4.37         $ $\pm$ $8.81\, 10^{-1}$ & $3.77\, 10^{2}$ $\pm$ $7.95\, 10^{1}$ \\
            & $1.19\, 10^{-1}$ & $4.18\, 10^{-1}$ & $2.40\, 10^{-1}$ & $2.23$ & $3.01\, 10^{1}$ $\pm$ $1.29\, 10^{ 1}$ & $5.05\, 10^{3}$ $\pm$ $2.19\, 10^{3}$ \\
\hline
EXO 0748    & $2.15\, 10^{-2}$ & $5.98\, 10^{-2}$ & $4.91\, 10^{-2}$ & $0.11$ & $8.85         $ $\pm$ $1.25          $ & $2.80\, 10^{2}$ $\pm$ $6.69\, 10^{1}$ \\
            & $5.98\, 10^{-2}$ & $7.87\, 10^{-2}$ & $6.90\, 10^{-2}$ & $0.43$ & $8.13         $ $\pm$ $1.15          $ & $2.93\, 10^{2}$ $\pm$ $6.97\, 10^{1}$ \\
            & $7.87\, 10^{-2}$ & $9.52\, 10^{-2}$ & $8.76\, 10^{-2}$ & $0.84$ & $5.71         $ $\pm$ $8.10\, 10^{-1}$ & $1.80\, 10^{2}$ $\pm$ $4.25\, 10^{1}$ \\
            & $9.52\, 10^{-2}$ & $1.09\, 10^{-1}$ & $1.03\, 10^{-1}$ & $1.06$ & $3.50         $ $\pm$ $4.95\, 10^{-1}$ & $1.79\, 10^{2}$ $\pm$ $4.26\, 10^{1}$ \\
            & $1.09\, 10^{-1}$ & $1.21\, 10^{-1}$ & $1.15\, 10^{-1}$ & $1.15$ & $2.46         $ $\pm$ $3.48\, 10^{-1}$ & $1.34\, 10^{2}$ $\pm$ $3.19\, 10^{1}$ \\
            & $1.21\, 10^{-1}$ & $1.38\, 10^{-1}$ & $1.27\, 10^{-1}$ & $1.17$ & $1.76         $ $\pm$ $2.52\, 10^{-1}$ & $1.16\, 10^{2}$ $\pm$ $2.78\, 10^{1}$ \\
            & $1.38\, 10^{-1}$ & $7.20\, 10^{-1}$ & $1.84\, 10^{-1}$ & $1.28$ & $4.78         $ $\pm$ $7.99\, 10^{-1}$ & $3.28\, 10^{2}$ $\pm$ $8.37\, 10^{1}$ \\
\hline
4U 1636     & $4.00\, 10^{-4}$ & $4.15\, 10^{-2}$ & $3.46\, 10^{-2}$ & $1.14$ & $3.04         $ $\pm$ $4.30\, 10^{-1}$ & $4.11\, 10^{1}$ $\pm$ $6.71         $ \\
            & $4.15\, 10^{-2}$ & $4.73\, 10^{-2}$ & $4.48\, 10^{-2}$ & $1.37$ & $2.56         $ $\pm$ $3.70\, 10^{-1}$ & $4.73\, 10^{1}$ $\pm$ $7.83         $ \\
            & $4.73\, 10^{-2}$ & $5.00\, 10^{-2}$ & $4.84\, 10^{-2}$ & $1.34$ & $1.65         $ $\pm$ $2.29\, 10^{-1}$ & $3.49\, 10^{1}$ $\pm$ $5.60         $ \\
            & $5.00\, 10^{-2}$ & $5.34\, 10^{-2}$ & $5.17\, 10^{-2}$ & $1.40$ & $2.02         $ $\pm$ $3.01\, 10^{-1}$ & $4.05\, 10^{1}$ $\pm$ $6.88         $ \\
            & $5.34\, 10^{-2}$ & $5.75\, 10^{-2}$ & $5.53\, 10^{-2}$ & $1.31$ & $2.58         $ $\pm$ $3.51\, 10^{-1}$ & $5.53\, 10^{1}$ $\pm$ $8.74         $ \\
            & $5.75\, 10^{-2}$ & $6.10\, 10^{-2}$ & $5.92\, 10^{-2}$ & $1.44$ & $2.05         $ $\pm$ $2.88\, 10^{-1}$ & $5.09\, 10^{1}$ $\pm$ $8.22         $ \\
            & $6.10\, 10^{-2}$ & $6.55\, 10^{-2}$ & $6.31\, 10^{-2}$ & $1.32$ & $2.18         $ $\pm$ $3.09\, 10^{-1}$ & $5.91\, 10^{1}$ $\pm$ $9.63         $ \\
            & $6.55\, 10^{-2}$ & $7.08\, 10^{-2}$ & $6.81\, 10^{-2}$ & $1.53$ & $2.17         $ $\pm$ $3.08\, 10^{-1}$ & $5.03\, 10^{1}$ $\pm$ $8.19         $ \\
            & $7.08\, 10^{-2}$ & $8.09\, 10^{-2}$ & $7.60\, 10^{-2}$ & $1.50$ & $3.50         $ $\pm$ $4.95\, 10^{-1}$ & $1.05\, 10^{2}$ $\pm$ $1.71\, 10^{1}$ \\
            & $8.09\, 10^{-2}$ & $9.72\, 10^{-2}$ & $8.98\, 10^{-2}$ & $1.52$ & $5.24         $ $\pm$ $7.40\, 10^{-1}$ & $1.57\, 10^{2}$ $\pm$ $2.57\, 10^{1}$ \\
            & $9.72\, 10^{-2}$ & $1.17\, 10^{-1}$ & $1.06\, 10^{-1}$ & $1.87$ & $6.17         $ $\pm$ $8.84\, 10^{-1}$ & $2.19\, 10^{2}$ $\pm$ $3.60\, 10^{1}$ \\
            & $1.17\, 10^{-1}$ & $1.51\, 10^{-1}$ & $1.31\, 10^{-1}$ & $2.10$ & $7.87         $ $\pm$ $1.10          $ & $3.85\, 10^{2}$ $\pm$ $6.27\, 10^{1}$ \\
            & $1.51\, 10^{-1}$ & $2.80\, 10^{-1}$ & $1.78\, 10^{-1}$ & $2.22$ & $1.84\, 10^{1}$ $\pm$ $4.53          $ & $1.74\, 10^{3}$ $\pm$ $4.55\, 10^{2}$ \\
\hline
4U 1608     & $3.20\, 10^{-3}$ & $1.56\, 10^{-2}$ & $1.00\, 10^{-2}$ & $0.89$ & $1.92\, 10^{1}$ $\pm$ $3.87          $ & $9.17\, 10^{1}$ $\pm$ $2.00\, 10^{1}$ \\
            & $1.56\, 10^{-2}$ & $2.44\, 10^{-2}$ & $1.99\, 10^{-2}$ & $1.03$ & $1.61\, 10^{1}$ $\pm$ $3.25          $ & $1.28\, 10^{2}$ $\pm$ $2.79\, 10^{1}$ \\
            & $2.44\, 10^{-2}$ & $3.28\, 10^{-2}$ & $2.82\, 10^{-2}$ & $1.20$ & $5.21         $ $\pm$ $1.05          $ & $6.42\, 10^{1}$ $\pm$ $1.40\, 10^{1}$ \\
            & $3.28\, 10^{-2}$ & $4.64\, 10^{-2}$ & $3.87\, 10^{-2}$ & $1.31$ & $2.43         $ $\pm$ $4.89\, 10^{-1}$ & $3.54\, 10^{1}$ $\pm$ $7.74         $ \\
            & $4.64\, 10^{-2}$ & $7.97\, 10^{-2}$ & $6.43\, 10^{-2}$ & $1.01$ & $3.10         $ $\pm$ $6.25\, 10^{-1}$ & $6.35\, 10^{1}$ $\pm$ $1.39\, 10^{1}$ \\
            & $7.97\, 10^{-2}$ & $3.92\, 10^{-1}$ & $1.72\, 10^{-1}$ & $1.76$ & $2.96\, 10^{1}$ $\pm$ $1.02\, 10^{ 1}$ & $7.79\, 10^{2}$ $\pm$ $2.76\, 10^{2}$ \\
\hline
\end{tabular}
\caption{\textbf{Data}. The first two columns indicate the range of each \minb{} bin, while the third and fourth columns are the $\mdo$ and $S_z$ position in the color-color diagram obtained as averages of the values in each bin. The last two columns are the average burst recurrence time
and $\alpha$ in each bin. As in \citet{rev-2020-gall-etal}, fluxes are corrected for anisotropy as $(\xi_{\rm{b}}, \xi_{\rm{p}}) = (0.898, 0.809)$ for sources without dips and $(\xi_{\rm{b}}, \xi_{\rm{p}}) = (1.639, 7.27)$ for sources with dips.}
\label{tab:data}
\end{table}

\section{Fit results}

\begin{table}
\centering
\begin{tabular}{l|lll|lll|}
\hline
Source & $\dot{m}_{\rm{loc}} / \dot{m}_{\rm{Edd}}$ & $\sigma^{-}_{\dot{m}_{\rm{loc}}} / \dot{m}_{\rm{Edd}}$ & $\sigma^{+}_{\dot{m}_{\rm{loc}}} / \dot{m}_{\rm{Edd}}$ & $w$ & $\sigma^{-}_{w}$ & $\sigma^{+}_{w}$ \\
\hline
KS 1731     & $5.00\, 10^{-2}$ & $5.18\, 10^{-3}$ & $8.99\, 10^{-3}$ & $9.08\, 10^{-1}$ & $6.74\, 10^{-2}$ & $6.30\, 10^{-2}$ \\
            & $1.39\, 10^{-1}$ & $2.48\, 10^{-2}$ & $9.32\, 10^{-3}$ & $1.00$ & $7.14\, 10^{-2}$ & $0.00$ \\
            & $1.15\, 10^{-1}$ & $2.57\, 10^{-2}$ & $1.62\, 10^{-3}$ & $1.00$ & $1.22\, 10^{-1}$ & $0.00$ \\
            & $1.65\, 10^{-1}$ & $3.29\, 10^{-2}$ & $1.37\, 10^{-2}$ & $1.00$ & $6.67\, 10^{-2}$ & $0.00$ \\
            & $1.52\, 10^{-1}$ & $3.54\, 10^{-2}$ & $6.34\, 10^{-3}$ & $1.00$ & $1.15\, 10^{-1}$ & $0.00$ \\
            & $1.06\, 10^{-1}$ & $1.87\, 10^{-2}$ & $1.02\, 10^{-2}$ & $9.23\, 10^{-1}$ & $1.06\, 10^{-1}$ & $4.58\, 10^{-2}$ \\
            & $4.66\, 10^{-2}$ & $3.11\, 10^{-3}$ & $4.74\, 10^{-3}$ & $3.38\, 10^{-1}$ & $2.34\, 10^{-2}$ & $1.75\, 10^{-2}$ \\
            & $5.37\, 10^{-2}$ & $7.11\, 10^{-3}$ & $2.60\, 10^{-2}$ & $1.44\, 10^{-1}$ & $1.09\, 10^{-1}$ & $6.05\, 10^{-2}$ \\
\hline
Aql X-1 & $3.96\, 10^{-2}$ & $2.68\, 10^{-3}$ & $3.91\, 10^{-3}$ & $2.92\, 10^{-1}$ & $4.31\, 10^{-2}$ & $3.14\, 10^{-2}$ \\
        & $8.94\, 10^{-2}$ & $1.82\, 10^{-2}$ & $1.13\, 10^{-2}$ & $2.79\, 10^{-1}$ & $2.46\, 10^{-2}$ & $7.56\, 10^{-2}$ \\
        & $6.68\, 10^{-2}$ & $8.42\, 10^{-3}$ & $1.67\, 10^{-2}$ & $3.05\, 10^{-1}$ & $5.41\, 10^{-2}$ & $3.75\, 10^{-2}$ \\
        & $6.26\, 10^{-2}$ & $5.03\, 10^{-2}$ & $0.00$ & $8.55\, 10^{-6}$ & $0.00$ & $5.65\, 10^{-1}$ \\
\hline
EXO 0748     & $4.01\, 10^{-2}$ & $6.27\, 10^{-3}$ & $2.06\, 10^{-3}$ & $3.73\, 10^{-1}$ & $1.11\, 10^{-2}$ & $1.87\, 10^{-1}$ \\
             & $4.19\, 10^{-2}$ & $6.39\, 10^{-3}$ & $2.71\, 10^{-3}$ & $3.69\, 10^{-1}$ & $1.44\, 10^{-2}$ & $1.60\, 10^{-1}$ \\
             & $4.91\, 10^{-2}$ & $9.75\, 10^{-3}$ & $1.16\, 10^{-3}$ & $3.95\, 10^{-1}$ & $0.00$ & $2.69\, 10^{-1}$ \\
             & $7.62\, 10^{-2}$ & $1.88\, 10^{-2}$ & $1.04\, 10^{-2}$ & $3.83\, 10^{-1}$ & $1.41\, 10^{-2}$ & $2.07\, 10^{-1}$ \\
             & $1.05\, 10^{-1}$ & $2.84\, 10^{-2}$ & $2.59\, 10^{-3}$ & $4.09\, 10^{-1}$ & $7.97\, 10^{-4}$ & $2.31\, 10^{-1}$ \\
             & $1.25\, 10^{-1}$ & $1.66\, 10^{-2}$ & $1.69\, 10^{-2}$ & $4.52\, 10^{-1}$ & $3.86\, 10^{-2}$ & $2.10\, 10^{-1}$ \\
             & $6.00\, 10^{-2}$ & $8.97\, 10^{-3}$ & $1.17\, 10^{-2}$ & $3.37\, 10^{-1}$ & $4.59\, 10^{-2}$ & $6.94\, 10^{-2}$ \\
\hline
4U 1636     & $1.14\, 10^{-1}$ & $2.05\, 10^{-2}$ & $4.05\, 10^{-3}$ & $1.00$ & $8.58\, 10^{-2}$ & $0.00$ \\
            & $1.26\, 10^{-1}$ & $2.90\, 10^{-2}$ & $2.51\, 10^{-3}$ & $9.95\, 10^{-1}$ & $1.41\, 10^{-1}$ & $0.00$ \\
            & $2.19\, 10^{-1}$ & $5.08\, 10^{-2}$ & $1.23\, 10^{-2}$ & $1.00$ & $5.98\, 10^{-2}$ & $0.00$ \\
            & $1.65\, 10^{-1}$ & $3.83\, 10^{-2}$ & $1.10\, 10^{-2}$ & $1.00$ & $9.53\, 10^{-2}$ & $0.00$ \\
            & $1.03\, 10^{-1}$ & $1.37\, 10^{-2}$ & $1.58\, 10^{-2}$ & $8.67\, 10^{-1}$ & $8.35\, 10^{-2}$ & $9.07\, 10^{-2}$ \\
            & $1.30\, 10^{-1}$ & $1.81\, 10^{-2}$ & $2.28\, 10^{-2}$ & $9.16\, 10^{-1}$ & $1.03\, 10^{-1}$ & $5.61\, 10^{-2}$ \\
            & $1.10\, 10^{-1}$ & $1.06\, 10^{-2}$ & $2.24\, 10^{-2}$ & $8.04\, 10^{-1}$ & $7.21\, 10^{-2}$ & $1.37\, 10^{-1}$ \\
            & $1.27\, 10^{-1}$ & $1.98\, 10^{-2}$ & $1.82\, 10^{-2}$ & $9.31\, 10^{-1}$ & $1.11\, 10^{-1}$ & $4.27\, 10^{-2}$ \\
            & $5.90\, 10^{-2}$ & $5.28\, 10^{-3}$ & $1.32\, 10^{-2}$ & $6.09\, 10^{-1}$ & $9.14\, 10^{-2}$ & $1.19\, 10^{-1}$ \\
            & $5.13\, 10^{-2}$ & $1.00\, 10^{-2}$ & $1.70\, 10^{-3}$ & $4.06\, 10^{-1}$ & $0.00$ & $1.79\, 10^{-1}$ \\
            & $4.75\, 10^{-2}$ & $7.15\, 10^{-3}$ & $3.74\, 10^{-3}$ & $3.84\, 10^{-1}$ & $6.99\, 10^{-3}$ & $9.91\, 10^{-2}$ \\
            & $4.43\, 10^{-2}$ & $2.80\, 10^{-3}$ & $4.05\, 10^{-3}$ & $3.43\, 10^{-1}$ & $2.12\, 10^{-2}$ & $1.64\, 10^{-2}$ \\
            & $7.18\, 10^{-2}$ & $2.56\, 10^{-2}$ & $0.00$ & $2.50\, 10^{-5}$ & $0.00$ & $1.40\, 10^{-1}$ \\
\hline
4U 1608     & $2.69\, 10^{-2}$ & $1.29\, 10^{-3}$ & $1.30\, 10^{-3}$ & $8.38\, 10^{-1}$ & $7.67\, 10^{-2}$ & $1.12\, 10^{-1}$ \\
            & $2.72\, 10^{-2}$ & $7.04\, 10^{-4}$ & $2.66\, 10^{-3}$ & $6.65\, 10^{-1}$ & $1.43\, 10^{-1}$ & $1.52\, 10^{-1}$ \\
            & $5.40\, 10^{-2}$ & $8.52\, 10^{-3}$ & $1.24\, 10^{-2}$ & $9.34\, 10^{-1}$ & $9.87\, 10^{-2}$ & $4.13\, 10^{-2}$ \\
            & $1.41\, 10^{-1}$ & $3.07\, 10^{-2}$ & $1.58\, 10^{-2}$ & $1.00$ & $8.44\, 10^{-2}$ & $0.00$ \\
            & $8.01\, 10^{-2}$ & $1.21\, 10^{-2}$ & $2.10\, 10^{-2}$ & $8.32\, 10^{-1}$ & $9.63\, 10^{-2}$ & $1.11\, 10^{-1}$ \\
            & $3.57\, 10^{-2}$ & $8.40\, 10^{-3}$ & $5.26\, 10^{-3}$ & $2.69\, 10^{-1}$ & $9.35\, 10^{-2}$ & $1.04\, 10^{-1}$ \\
\hline
\end{tabular}
\caption{Theoretical fitted local mass accretion rate and composition purity. Errors correspond to 1 $\sigma$ (16 and 84 \%).}
\label{tab:fitres}
\end{table}

\resufigM{}
\resufigS{}

Here, we report the results of the fits.
We also report on a curious finding of our fits. In \fig{fig:peaks} we show the $\mdt$ at minimum burst recurrence time, which resulted from our fits to each source, together with their weighted average (0.13, corrected for asymmetry of the error bars \citealt{rep-2000-schme}). It is interesting to note that they seem vaguely compatible with being the same (although the reduced $\chi^2$ is $\sim 4$). This value could represent some kind of threshold in the fluid dynamics, triggering the onset of instabilities in the spreading layer and affecting mixing of the fuel. From the point of view of accretion, since this threshold $\mdt$ needs to correspond to a global mass accretion rate which decreases with spin, it could imply that the area covered by the (hot) flow at this point scales inversely with the spin of the star, which could again be compatible with the effect of a stronger Coriolis force. Note, however, that the values in \fig{fig:peaks} are affected by the choices of the bins in the data; in particular, they could be higher if smaller bins show higher peak rates, so the analyses should be redone when more resolution is possible, before drawing strong conclusions. Also note that the existence of a giant solitary wave was speculated to be necessary to make a spreading layer compatible with the presence of the bursts \citep{art-2010-inoga-suny}. Without the wave, the spreading layer would liberate too much heat in the deep layers and quench the bursts. It would also be too thick to develop within observed time scales. It is possible that the $\mdt$ we find corresponds to a value above which accretion begins to be so high that it interferes with this wave, thus leading to a hotter column and, in turn, helping the generation of heavier elements and the quenching of the bursts.

\figmax{}

\end{document}